\documentclass[12pt]{article}
\newcommand{\beq}{\begin{equation}}
\newcommand{\eeq}{\end{equation}}
\newcommand{\bdis}{\begin{displaymath}}
\newcommand{\edis}{\end{displaymath}}
\newcommand{\beqa}{\begin{eqnarray}}
\newcommand{\beas}{\begin{eqnarray*}}
\newcommand{\eeqa}{\end{eqnarray}}
\newcommand{\eeas}{\end{eqnarray*}}
\newcommand{\ba}{\begin{array}}
\newcommand{\ea}{\end{array}}
\newcommand{\bea}{\begin{eqnarray}}
\newcommand{\eea}{\end{eqnarray}}

\usepackage{graphicx}
\begin{document}

\title{Exactly solvable $N$-body  quantum systems with $N=3^k \ ( k \geq 2)$ in the $D=1$ dimensional space}
\author{ A. Bachkhaznadji \\
Laboratoire de Physique Th\'eorique,
D\'epartement de Physique \\
Universit\'e  Mentouri,
Constantine, Algeria\\ 
\\
M. Lassaut   \\
Institut de Physique Nucl\'eaire, CNRS-IN2P3, Universit\'e Paris-Sud, \\
 Universit\'e   Paris-Saclay,   F-91406 Orsay Cedex,  France \\ [3mm]
\date{\today }}
\maketitle

\noindent
Abstract :

\vspace{0.5cm}
\noindent
We study the exact solutions of a particular class of $N$ confined  particles of equal mass, with $N=3^k \ (k=2,3,...),$ in the $D=1$ dimensional space.  The particles are clustered in clusters of 3 particles.
 The interactions involve a confining mean field, two-body Calogero type of potentials inside the cluster, 
  interactions between the  centres of mass of the clusters and finally a non-translationally invariant 
  $N$-body potential. The case of 9 particles is exactly solved, in a first step, by providing the full eigensolutions and eigenenergies.  Extending this procedure, the general case of $N$ particles ($N=3^k, \ k \geq 2$)  is studied in a second step. 
The exact solutions are obtained via appropriate coordinate transformations and separation of variables.
The eigenwave functions and the corresponding energy spectrum are provided.

\vspace{2cm}
\newpage
\pagestyle{plain}
\pagenumbering{arabic}
\setcounter{page}{2}
\baselineskip=14pt
\vspace{-2cm}
\section{Introduction}

The study of exactly solvable integrable quantum systems of $N$ interacting particles still retains attention.
The Calogero \cite{Calo69p,Calo71} and Sutherland \cite{Suth71,Suth71bis} models constitute  famous examples.
Most of the works have been performed in the $D=1$ dimensional space.
A good survey of the many-body problems  can be found in  \cite{mattis,sutherland}, and in the report 
\cite{Perelomov1983}, where many integrable quantum systems have been classified with respect to Lie algebras.
 Systems with point interactions have also been considered, still in $D=1$  \cite{albe1,albe2}.
The early works of Calogero and Sutherland have been extended to many more complex systems, concerning two- and three-body problems.  We can quote, in a non exhaustive way, the works of several authors [10-19].

For four-body systems and beyond, the works on exactly solvable quantum systems 
 are much more scarce [20-23].
 Recently we have  solved exactly some few-body quantum problems consisting of four, five and six particles moving on a  line \cite{BL15}. In these particular systems, the particles are confined in a harmonic trap and interact pairwise, in clusters of two and three particles, through two-body inverse square Calogero potentials. The obtained results suggest to extend the construction  to larger
 systems of particles, even for $N$-body problems, with large values of $N$.

 The purpose of the present paper is to point out a particular case of  a $N$-body quantum problem admitting an exact solution. The interactions between the particles of this system are inspired by those  used in our previous works
  for 3,4,5 and 6-particle systems \cite{BLL2009,BL13,BL15}. The $N$ particles, of equal masses, are confined in a harmonic mean field with frequency $\omega$, and clustered in clusters of three particles. Then, the number $N$ of  particles is  $N=3^k \ (k=2,3,...).$ 
 In each of $3^{k-1}$ three-body clusters, the particles interact mutually with two-body inverse square Calogero potentials  
\cite{Calo71}.
 Other  many-particle interactions are added and  chosen as follows : the clusters interact via a Calogero type of potential 
depending on the distance between their centres of mass. This can be generalized to $N=3^k$ particles with $ k \geq 2$ by clustering again the clusters in groups of 3 clusters, and letting the clusters of clusters to interact via their centres of mass.
Finally a non-translationally invariant $N$-body interaction is added.

The hierarchy  of clustering is illustrated in Fig.{\bf 1}  for  the case of $N= 3^2$ particles.
\begin{figure}[h]
\includegraphics[scale=0.5]{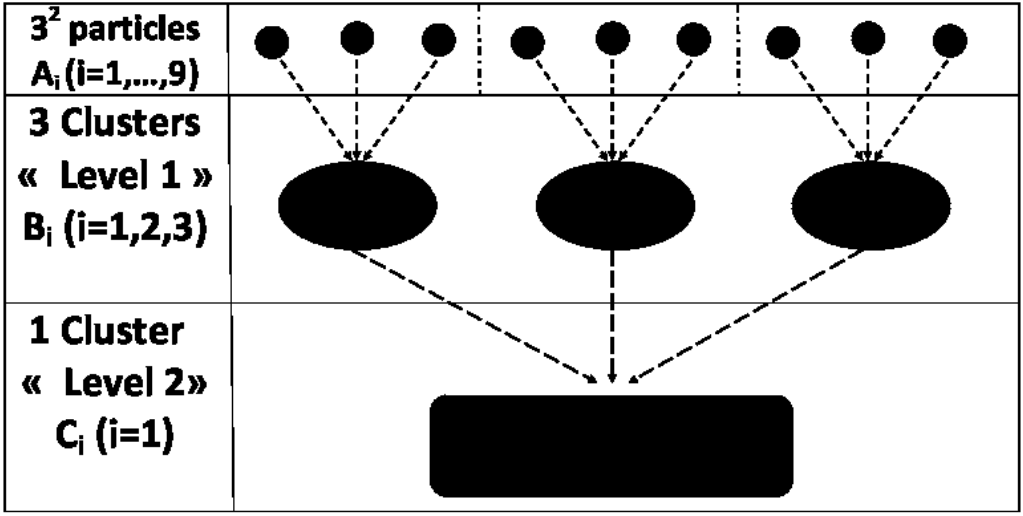}
\caption{\label{f1} {\bf  A)} $3^2$ particles in interaction,  {\bf B)} : 3 clusters of "level 1" each containing 3 particles,  {\bf C)}  one  cluster of "level 2"  containing 3 clusters of  "level 1".}
\end{figure}
The case of 9 particles is studied and solved exactly. 
 With some  successive and appropriate coordinate transformations, the problem is solved by separation of variables.  The solutions of the corresponding stationary Schr\"odinger equation are provided, namely the eigenwavefunctions and the corresponding eigenenergies.
The general case of $N=3^k, \ (k \geq 2)$ is then studied by following the same procedure.
 After some straightforward calculations  and coordinates transformations, the exact solutions are given.

The paper is organized   as follows.
In section {\bf 2} we present and solve the nine-body problem
 for the case of harmonic confinement of the particles.
 In section {\bf 3} we treat the case of $N=3^k$ particles. 
 Conclusions are given  in section {\bf 4}.

\section{A nine-body  problem  with  harmonic confinement}

 We consider the Hamiltonian : 

\begin{eqnarray}
H &= &\sum_{i=1}^{9}\left( -\frac{\partial ^{2}}{\partial x_{i}^{2}}%
+\omega ^{2}x_{i}^{2}\right) + \sum_{1 \leq i<j \leq 3} \frac{\lambda_{1,1}}{(x_{i}-x_{j})^{2}}+ \sum_{4 \leq i<j \leq 6}  \frac{\lambda_{2,1}}{(x_{i}-x_{j})^{2}}
  +  \sum_{7 \leq i<j  \leq 9} \frac{\lambda_{3,1}}{(x_{i}-x_{j})^{2}}  \nonumber\\ 
& + &   \sum_{1 \leq i<j \leq 3} \frac{3 \lambda_{1,2}}{  \left( x_{3 i-2}+x_{3 i-1} + x_{3 i} -
x_{3 j -2}-x_{3 j-1} - x_{3 j}  \right)^2}
 +    \frac{\mu}{ \sum_{i=1}^{9} x_i^2} 
\label{3cb-11}
\end{eqnarray}

Here, we use the units $\hbar=2 m =1$. The first term gives the energy of the nine independent particles 
with coordinates $x_i,i=1,2,..,9$ in a harmonic trap. The nine particles are clustered in 3 clusters  of 3 particles.
  Inside each cluster, the particles     interact pairwise  via a two-body inverse square  Calogero potential. 
  The first cluster involves the  first three particles, with coordinates $x_1,x_2$ and $x_3$,
  the second one the next three   particles, with coordinates $x_4,x_5$ and $x_6$,
  and the third one the   particles  with coordinates $x_7,x_8$ and $x_9$.
   The next-to-last  terms represent the  3  clusters interacting  pairwise via their centre of mass. 
   It gives rise to the terms $ (x_{3 i-2}+x_{3 i-1} + x_{3 i} - x_{3 j -2}-x_{3 j-1} - x_{3 j})^-2$  
  and  constitutes actually a 6-body interaction.     
 A non-translationally invariant nine-body potential with coupling constant $\mu$ is added, represented by the last term 
$ \mu/(\sum_{i=1}^{9} x_i^2)$. 
 
In order to solve this nine-body problem,
let us introduce the  first coordinate transformation to Jacobi and centre of mass  coordinates 

\begin{eqnarray}
&& u_{i,1}=\frac{1}{\sqrt{2}}(x_{3 i -2}-x_{3 i -1}),   v_{i,1} = \frac{1}{\sqrt{6}}(x_{3 i -2}+x_{3 i -1}-2x_{3 i}) 
 \label{3cb-16bis} \\
& &   w_{i,1}= \frac{1}{\sqrt{3}} (x_{3 i-2}+x_{3 i-1} + x_{3 i}) \qquad\quad
 i=1,2,3  \ .
\end{eqnarray}
The second index "1" refers to this first coordinate transformation.
The transformed Hamiltonian reads:
\begin{eqnarray}
H &= &\sum_{i=1}^{3} \left[ -\frac{\partial ^{2}}{\partial u_{i,1}^2}
-\frac{\partial^{2}}{\partial v_{i,1}^2}  -\frac{\partial ^{2}}{\partial w_{i,1}^{2}}+
 \omega^2 [ u_{i,1}^2+v_{i,1}^2+w_{i,1}^2] +
 \frac{9\lambda_{i,1} [u_{i,1}^2+v_{i,1}^2]^2}{2\left[u_{i,1}^3 -   3 u_{i,1} v_{i,1}^2 \right]^{2}}  \right]  \nonumber\\  
 & +  &   \sum_{1 \leq i<j \leq 3} \frac{\lambda_{1,2}}{(w_{i,1}- w_{j,1})^2 }     + \frac{\mu}{\sum_{i=1}^3 [u_{i,1}^2+v_{i,1}^2+w_{i,1}^2]}  \  .
\label{3cb-48}
\end{eqnarray}

This Hamiltonian is not  separable in the 9 variables  $\{u_{i,1},v_{i,1},w_{i,1}\}, i=1,2,3$.
We first introduce  polar  coordinates  : 
\begin{equation}
 u_{i,1} = r_{i,1}  \sin \varphi_{i,1},\quad v_{i,1}=r_{i,1} \cos \varphi_{i,1} ,\quad
 0 \leq r_{i,1}<\infty, \quad 0\leq \varphi_{i,1} \leq 2\pi, \ \quad i=1,2,3  \ .
\label{3cb-18bis}
\end{equation}
By using  the notations
\begin{eqnarray}
\tilde w_{1} & \equiv & \{w_{1,1},w_{2,1},w_{3,1} \},\tilde r_{1}  \equiv  \{r_{1,1},r_{2,1}, r_{3,1} \},
\tilde \varphi_{1}  \equiv  \{\varphi_{1,1},\varphi_{2,1}, \varphi_{3,1} \},
\end{eqnarray} 
 the Schr\"{o}dinger equation is then written as :
\begin{eqnarray}
&&\left\{ \sum_{i=1}^3  \left(-\frac{\partial ^{2}}{\partial w_{i,1}^2} -\frac{\partial ^{2}}{\partial r_{i,1}^2}
-\frac{1}{r_{i,1}}\frac{\partial }{\partial r_{i,1} }+  \omega ^{2} [r_{i,1}^2+w_{i,1}^2]  \right) \right. \nonumber\\
& & \left. + 
   \sum_{1 \leq i<j \leq 3} \frac{\lambda_{1,2}}{(w_{i,1}- w_{j,1})^2 }  
 +\frac{\mu}{\sum_{i=1}^3 (r_{i,1}^2 +w_{i,1}^2 )}  \right. \nonumber\\
   & & \left. + \sum_{i=1}^3 \frac{1}{r_{i,1}^2} \left[-\frac{\partial ^{2}}{\partial \varphi_{i,1}^2}+\frac{9\lambda_{i,1} }{2\sin^{2}(3\varphi_{i,1} )} \right] -E    \right\} \Psi(\tilde w_1,\tilde r_1,\tilde \varphi_1) =0  .  \label{3cb-49}
\end{eqnarray}%


The potential involved in the equation (\ref{3cb-49})
\begin{eqnarray}
V(\tilde w_1,\tilde r_1,\tilde \varphi_1)&= &
\sum_{1 \leq i<j \leq 3} \frac{\lambda_{1,2}}{(w_{i,1}- w_{j,1})^2 }+\frac{\mu}{\sum_{i=1}^3 (r_{i,1}^2+w_{i,1}^2)}
 + \sum_{i=1}^3  \left[\frac{1}{r_{i,1}^{2}}\  \frac{9\lambda_{i,1}}{2\sin^{2}(3\varphi_{i,1} )} \right] \nonumber
\end{eqnarray}

 has the general form 
\begin{equation}
V(\tilde w_1,\tilde r_1,\tilde \varphi_1)=f(\tilde w_1,\tilde r_1)+\sum_{i=1}^3  \frac{f_i(\varphi_{i,1})}{r_{i,1}^2} \ .
\end{equation}
This suggests the wave function to be factorized as follows
\begin{equation}
\Psi(\tilde w_1,\tilde r_1,\tilde \varphi_1) = \chi(\tilde w_1,\tilde r_1) \times \prod_{i=1}^3
 \Phi_{i,1}(\varphi_{i,1} )  \ .
\end{equation}
The equation (\ref{3cb-49})  will be solved  in two steps.
Firstly we consider  the 3 angular equations : 
\begin{equation}
\left( -\frac{d^{2}}{d\varphi_{i,1}^{2}}+\frac{9\lambda_{i,1} }{2\sin^{2}(3\varphi_{i,1}) }%
\right) \Phi_{n_{i,1}}(\varphi_{i,1} )=B_{n_{i,1}} \Phi_{n_{i,1}}(\varphi_{i,1} ), \qquad\quad  i=1,2,3 \ , \quad  \label{3cb1-7}
\end{equation}
on the interval $]0,\pi/3[$,  with Dirichlet conditions at the boundaries.
In  the vicinity of  $\varphi_{i,1}=0, (i=1,2,3)$ (resp. $\frac{\pi }{3},$) the singularity
can be treated if and only if $\lambda_{i,1} >-1/2$, similar to the case
 of a centrifugal barrier.  Otherwise the operator has several self-adjoint extensions, 
each of which may lead to a different spectrum \cite{Zno,Barry}.

The $B_{n_{i,1}}, i=1,2,3,$  are the  eigenvalues of  the equations (\ref{3cb1-7}), 
respectively  given by \cite{Calo69,BLL2009}
\begin{eqnarray}
& & B_{n_{i,1}} = b_{n_{i,1}}^2, \qquad b_{n_{i,1}}=3\left( n_{i,1} + \frac{1}{2} + a_{i,1} \right),\label{3cb-22-bis} \\
& &   a_{i,1} =\frac{1}{2} \sqrt{1+2\lambda_{i,1} }, \left(\lambda_{i,1} > -\frac{1}{2} \right),  \ \qquad n_{i,1}=0,1,2,.... , \quad i=1,2,3  \ . \label{3cb-22-ter}
\end{eqnarray}
The associated eigensolutions are given in terms of the Gegenbauer polynomials $C_n^{(q)}$ \cite{Abramowitzbook}
\begin{eqnarray}
\Phi_{n_{i,1}}(\varphi_{i,1} ) & = & (\sin 3\varphi_{i,1} )^{\frac{1}{2}+ a_{i,1}} C_{n_{i,1}}^{(\frac{1}{2}+
a_{i,1})}(\cos 3\varphi_{i,1} ),   \label{3cb-22}  \\
& &   \quad 0\leq \varphi_{i,1} \leq \frac{\pi }{3},    
          \qquad n_{i,1}=0,1,2,.. \ . \nonumber
\end{eqnarray}

The extension from the interval $]0,\pi/3[$ to the whole interval $[0,2 \pi]$ is made following the prescription given in 
\cite{Calo69} by using symmetry arguments according to the statistics obeyed by the particles.

The second step consists in the resolution of the  following Schr\"{o}dinger equation :

\begin{eqnarray}
& &\left\{ \sum_{i=1}^3 \left[-\frac{\partial ^{2}}{\partial w_{i,1}^{2}} -\frac{\partial ^{2}}{\partial r_{i,1}^{2}}-\frac{1}{r_{i,1}}\frac{\partial }{\partial r_{i,1}}+  \omega ^{2} 
[r_{i,1}^{2}+w_{i,1}^2] \right]  +  \sum_{1 \leq i<j \leq 3} \frac{\lambda_{1,2}}{(w_{i,1}-w_{j,1})^{2}}  \right.\nonumber\\ 
&  & \left. + \frac{\mu}{\sum_{i=1}^3   [r_{i,1}^{2}+w_{i,1}^2] } 
+ \sum_{i=1}^3 \frac{B_{n_{i,1}}}{r_{i,1}^2}   -E_{\tilde n_1}    \right\}
 \chi_{\tilde n_1}(\tilde w_1,\tilde r_1)
  =0  \ ,  \label{3cbb-49}
\end{eqnarray}%
with $\tilde n_1 \equiv \{n_{1,1},n_{2,1}, n_{3,1}\}$.

We first draw the  attention to $w_{i,1}$'s variables and  introduce the coordinates labeled by the second index  "2", corresponding to the second set of the coordinates transformation to the Jacobi and centre of mass coordinates
\begin{equation}
 u_{1,2}=\frac{1}{\sqrt{2}}(w_{1,1}-w_{2,1}),  v_{1,2}  = \frac{1}{\sqrt{6}}(w_{1,1}+w_{2,1}-2w_{3,1}) ,
  w_{1,2} = \frac{1}{\sqrt{3}} (w_{1,1}+w_{2,1} + w_{3,1}) \ .
\label{3c-16ter}
\end{equation}
 The transformed equation reads:
\begin{eqnarray}
 & & \left\{-\frac{\partial ^{2}}{\partial u_{1,2}^{2}}-\frac{\partial ^{2}}{\partial v_{1,2}^{2}}-
 \frac{\partial ^{2}}{\partial w_{1,2}^{2} } -
 \sum_{i=1}^3 \left(\frac{\partial ^{2}}{\partial r_{i,1}^{2}}+\frac{1}{r_{i,1}}\frac{\partial }{\partial r_{i,1}} \right) + \sum_{i=1}^3 \frac{B_{n_{i,1}}}{r_{i,1}^{2}}  \right. \nonumber\\
 & & \left. +\omega^{2}[ u_{1,2}^{2}+ v_{1,2}^{2}+ w_{1,2}^{2} +
 \sum_{i=1}^3 r_{i,1}^{2} ]  +
\frac{\mu }{ u_{1,2}^{2}+ v_{1,2}^{2}+ w_{1,2}^{2} + \sum_{i=1}^3 r_{i,1}^{2}  } \right. \nonumber \\ 
& &\left. + \frac{9 \lambda_{1,2}[ u_{1,2}^{2}+ v_{1,2}^{2}]^2}{2\left[u_{1,2}^{3}- 3 u_{1,2} v_{1,2}^{2} \right]^{2}}     -E_{\tilde n_1 } \right\} \chi_{\tilde n_1}(u_{1,2},v_{1,2},w_{1,2},\tilde r_1) =0 \ .\label{3c-48}
\end{eqnarray}

We then introduce the following  transformation to  polar  coordinates : 
\begin{eqnarray}
 u_{1,2} &= &r_{1,2}  \sin \varphi_{1,2},\quad v_{1,2}=r_{1,2} \cos \varphi_{1,2} ,\quad
 0 \leq r_{1,2}<\infty, \quad 0\leq \varphi_{1,2} \leq 2\pi \ .
\label{3cb-18ter}
\end{eqnarray}

The Schr\"{o}dinger equation(\ref{3c-48})  becomes : 
\begin{eqnarray}
&& \left\{   -\frac{\partial ^{2}}{\partial w_{1,2}^2}  -\frac{\partial ^{2}}{\partial r_{1,2}^2}-\frac{1}{r_{1,2}} \frac{\partial }{\partial r_{1,2}} -  \sum_{i=1}^3 \left( \frac{\partial ^{2}}{\partial r_{i,1}^{2}}+\frac{1}{r_{i,1}}\frac{\partial }{\partial r_{i,1}} \right)
+\omega ^{2} [w_{1,2}^{2}+ r_{1,2}^{2}+\sum_{i=1}^3 r_{i,1}^{2} 
  ]\right.  \nonumber\\
& & \left. +\frac{\mu }{w_{1,2}^{2}+ r_{1,2}^{2}+\sum_{i=1}^3 r_{i,1}^{2} }
 +  \frac{1}{r_{1,2}^{2}}\left[-\frac{\partial^2 }{\partial \varphi_{1,2}^2}
 + \frac{9 \lambda_{1,2}}{2 \sin^2 (3 \varphi_{1,2} )} \right] \right. \nonumber\\
 & & \left. +  \sum_{i=1}^3 \frac{B_{n_{i,1}}}{r_{i,1}^{2}}-E_{\tilde n_1}   \right\} 
 \chi_{\tilde n_1}(w_{1,2},r_{1,2}, \varphi_{1,2},\tilde r_1) =0 \ .
\label{3c-49}
\end{eqnarray}%

The potential involved in the equation (\ref{3c-49})
\begin{eqnarray}
V(w_{1,2},r_{1,2}, \varphi_{1,2},\tilde r_1)  &= &  \omega ^{2} [w_{1,2}^{2}+ r_{1,2}^{2}+\sum_{i=1}^3 r_{i,1}^{2}]+\frac{\mu }{w_{1,2}^{2}+ r_{1,2}^{2}+\sum_{i=1}^3 r_{i,1}^{2} } \nonumber\\
&+ &  \sum_{i=1}^3 \frac{B_{n_{i,1}}}{r_{i,1}^{2}} + \frac{1}{r_{1,2}^{2}}\  \frac{9 \lambda_{1,2}}{2\sin^{2}(3\varphi_{1,2})} \ ,
\label{potcp}
\end{eqnarray}
 has the general form 
\begin{equation}
V(w_{1,2},r_{1,2}, \varphi_{1,2},\tilde r_1)=f_1(w_{1,2},r_{1,2},\tilde r_1)+
 \frac{f(\varphi_{1,2})}{r_{1,2}^{2}} \ .
\end{equation}
Again, it suggests the wave function $ \chi_{\tilde n_1}$ to be factorized as follows :
\begin{equation}
 \chi_{\tilde n_{1}}(w_{1,2},r_{1,2}, \varphi_{1,2},\tilde r_1)
 = \frac {1}{\sqrt{r_{1,2} \prod_{i=1}^3 r_{i,1}}}
\eta_{\tilde n_1}(w_{1,2},r_{1,2}, \tilde r_1)  \  \Phi_{1,2}(\varphi_{1,2})  \ .
\end{equation}
The equation (\ref{3c-49})  will be solved  in two steps.
Firstly we solve 
\begin{equation}
\left( -\frac{d^{2}}{d \varphi_{1,2}^2}+\frac{9 \lambda_{1,2}  }{2\sin^{2}(3\varphi_{1,2})}%
\right) \Phi_{n_{1,2}}(\varphi_{1,2})=B_{n_{1,2}} \Phi_{n_{1,2} } (\varphi_{1,2} ) ,\quad  \label{3cb4-7}
\end{equation}
on the interval $]0,\pi/3[$,  with Dirichlet conditions at the boundaries.
Note that the condition $ \lambda_{1,2} > -\frac{1}{2}$ ensures that the operator is self-adjoint.
 The latter equation shows that 
$\Phi_{n_{1,2}}$ does not depend on the index $\tilde n_1$.

$B_{n_{1,2}}$  denotes  the  eigenvalues of  Eq.(\ref{3cb4-7}), given by 
\begin{eqnarray}
B_{n_{1,2}} & =& b_{n_{1,2}}^2 \qquad b_{n_{1,2}}=
3\left( n_{1,2} + \frac{1}{2} + a_{1,2} \right),  \label{3cbb-22-bis} \\
 a_{1,2} &=&\frac{1}{2} \sqrt{1+2 \lambda_{1,2}} \quad \left( \lambda_{1,2} > -\frac{1}{2}\right) 
, \qquad\quad  n_{1,2}=0,1,2,.... \ .   \ . \label{3cbb-22-ter}
\end{eqnarray}

The associated eigensolutions are written  in terms of the Gegenbauer polynomials $C_M^{(q)}$ 
\begin{eqnarray}
 \Phi_{n_{1,2}} (\varphi_{1,2} ) &= & [\sin 3 (\varphi_{1,2})]^{\frac{1}{2}+ a_{1,2}} \  C_{n_{1,2}}^{(\frac{1}{2}+a_{1,2})}(\cos 3 \varphi_{1,2} ),  \label{3cbb-22} \\
 & &    \quad 0\leq  \varphi_{1,2} \leq \frac{\pi }{3},    
          \qquad  n_{1,2}=0,1,2,...  \ .  \nonumber 
\end{eqnarray}

Then, we have to solve the  following Schr\"{o}dinger equation 

\begin{eqnarray}
 & & \left\{ - \frac{\partial ^{2}}{\partial w_{1,2}^{2} } - \frac{\partial ^{2}}{\partial r_{1,2}^{2} }- \sum_{i=1}^3 \left[\frac{\partial ^{2}}{\partial r_{i,1}^{2}} \right]
+\omega^{2}[ r_{1,2}^{2}+ w_{1,2}^{2} +
 \sum_{i=1}^3 r_{i,1}^{2} ] + \frac{\mu }{r_{1,2}^{2}+ w_{1,2}^{2} + \sum_{i=1}^3 r_{i,1}^{2}} \right. \nonumber\\
 & & \left. + \sum_{i=1}^3 \frac{B_{n_{i,1}}-\frac{1}{4}}{r_{i,1}^{2}} +
\frac{B_{n_{1,2}}-\frac{1}{4}}{r_{1,2}^{2}}
-E_{n_{1,2},\tilde n_1 } \right\} \eta_{n_{1,2},\tilde n_1}(w_{1,2},r_{1,2},\tilde r_1) 
=0 \ .
\label{3cc-49}
\end{eqnarray}%
For $\mu=0$, the solution of Eq.(\ref{3cc-49}) is simply given  in terms of the Laguerre Polynomials $L_k^{\ell}$
 and Hermite polynomials $H_M$  \cite{Abramowitzbook} :

\begin{eqnarray}
& &  \eta_{M,k_{1,2}, \tilde k_1, n_{1,2},\tilde n_1}(w_{1,2},r_{1,2},\tilde r_1)= H_{M}( \sqrt{\omega} w_{1,2}) \ \exp(-\omega w_{1,2}^2/2)  \nonumber\\
& & r_{1,2}^{ b_{n_{1,2}}+1/2} \ L_{k_{1,2}}^{b_{n_{1,2}}}(\omega r_{1,2}^{2})
\exp(-\omega  r_{1,2}^{2}/2)   \prod_{i=1}^3 r_{i,1}^{ b_{n_{i,1}}+1/2} L_{k_{i,1}}^{b_{n_{i,1}}}(\omega r_{i,1}^2) \exp(-\omega  r_{i,1}^{2}/2) \nonumber\\
& & M=0,1,2,... \ , \quad (\forall i), i=1,2,3,  k_{i,1}=0,1,2,...  \ , \quad \quad k_{1,2}=0,1,2,...  \  ,
\end{eqnarray}
and corresponds to the energy spectrum 
\begin{equation}
E_{M,k_{1,2}, \tilde k_1,n_{1,2},\tilde n_1 } = \omega \left( 9 + 2 b_{n_{1,2}} + 2 \sum_{i=1}^3  b_{n_{i,1}} + 
  2 M + 4  k_{n_{1,2}} + 4 \sum_{i=1}^3  k_{n_{i,1}} \right)  \  . \label{enercart}
\end{equation}
The energy (\ref{enercart}) can be rewritten, thanks to Eqs.(\ref{3cb-22-bis},\ref{3cbb-22-bis})
\begin{equation}
E_{M,k_{1,2}, \tilde k_1,n_{1,2},\tilde n_1 }=2 \omega \left\{ \frac{21}{2}  + 3 n_{1,2} + 3 a_{1,2}
+   \sum_{i=1}^3   (3 n_{i,1} + 3 a_{i,1}) + M + 2  k_{n_{1,2}} + 2 \sum_{i=1}^3  k_{n_{i,1}} \right \}   \ .
\label{enercartp}
\end{equation}

For  $\mu \ne 0$,  we introduce the hyperspherical transformation \cite{erd} :
\begin{eqnarray}
w_{1,2} &=& r \cos \alpha,  \quad\quad\quad\quad r_{1,2} = r \sin \alpha \cos \theta \nonumber\\
     0 & \leq &  r <\infty      \quad\quad\quad\quad\quad  0 \leq \alpha \leq \pi  \nonumber\\
r_{1,1} & = & r \sin \alpha \sin \theta \cos \beta, \quad r_{2,1}=r \sin \alpha \sin \theta \sin \beta  \sin \phi,
\quad r_{3,1}=r \sin \alpha \sin \theta \sin \beta \cos \phi ,\quad  \nonumber \\
  0 & \leq &  \theta \leq \frac{\pi}{2}, \quad\quad \quad\quad\quad\quad\quad  0\leq \beta \leq \frac{\pi}{2}, \quad \quad\quad\quad\quad\quad\quad\quad  0\leq \phi \leq \frac{\pi}{2}.  \label{3ca-18bis}
\end{eqnarray}

The Schr\"{o}dinger equation (\ref{3cc-49})  is then written as:
\begin{eqnarray}
&&\left\{ -\frac{\partial ^{2}}{\partial r^{2}}-\frac{4}{r}\frac{\partial }{%
\partial r}+\omega ^{2}r^{2}+\frac{\mu }{r^{2}}  +  \frac{1}{r^{2}}\left[
 -\frac{\partial ^{2}}{\partial \alpha ^{2}}   -3 \cot \alpha \frac{\partial }{\partial \alpha } + 
\frac{1}{\sin ^{2}\alpha}  \left(  -\frac{\partial ^{2}}{\partial \theta ^{2}}- 2 \cot \theta \frac{\partial }{\partial \theta }
 \right.\right. \right. \nonumber\\
& &\left.\left.\left. +  \frac{B_{n_{1,2}}-\frac{1}{4}}{ \cos ^{2}\theta } 
 + \frac{1}{\sin^2 \theta}  \left(   -\frac{\partial ^{2}}{\partial \beta ^{2}}-  \cot \beta \frac{\partial }{\partial \beta } + \frac{B_{n_{1,1}}-\frac{1}{4}}{ \cos^2 \beta} + \frac{1}{\sin ^{2}\beta }\left(- 
\frac{\partial ^{2}}{\partial \phi ^{2}} \right.\right.\right.\right.\right. \nonumber\\
 & & \left.\left.\left.\left.\left.+  \frac{B_{n_{2,1}}-\frac{1}{4}}{ \sin^2 \phi} +  \frac{B_{n_{3,1}}-\frac{1}{4}}{ \cos^2 \phi}\right) \right) \right) \right]   -E_{ n_{1,2},\tilde n_1 } \right\} \eta_{ n_{1,2},\tilde n_1}
(r,\alpha,\theta,\beta,\phi) =0  \ .  \label{3ca-49}
\end{eqnarray}%

This Hamiltonian may be mapped to
 the problem of one particle in the five dimensional space with a non central potential of the form
\begin{equation}
V(r,\alpha,\beta,\theta ,\phi )=f_{1}(r)+\frac{1}{r^{2}\sin ^{2}\alpha } \left[
f_{2}(\theta )+\frac{1}{\sin^2 \theta}  \left( f_{3}(\beta  ) +  \frac{f_{4}(\phi )}{\sin ^{2}\beta } \right)  \right].
\end{equation}
The problem becomes then  separable  in the five variables $\{r,\alpha,\theta,\beta,\varphi\}$.
To find the solution we factorize the wave function as follows :
\begin{equation}
\eta_{k,\ell,j,m,i, n_{1,2},\tilde n_1 } (r,\alpha,\theta ,\beta,\phi )=
\frac{F_{k,\ell,j,m,i,n_{1,2},\tilde n_1  }(r)}{r^2} \frac{G_{\ell,j,m,i,n_{1,2},\tilde n_1}(\alpha)}{\sin^{3/2} \alpha} \frac{\Theta_{j,m,i,n_{1,2},\tilde n_1} (\theta )}{\sin\theta } \frac{Q_{m,i,\tilde n_1}(\beta)}{\sqrt{\sin \beta}}
\zeta_{i,n_{2,1},n_{3,1}}(\phi ).  \nonumber
\end{equation}
Accordingly,  Eq.(\ref{3ca-49}) separates in  five   decoupled  differential equations:
\begin{equation}
\left( -\frac{d^{2}}{d\phi ^{2}}+ \frac{B_{n_{2,1}}-\frac{1}{4}}{ \sin^2 \phi}
+  \frac{B_{n_{3,1}}-\frac{1}{4}}{ \cos^2 \phi}\right) 
 \zeta_{i,n_{2,1},n_{3,1}}(\phi)   =R_{i,n_{2,1},n_{3,1}} \zeta_{i,n_{2,1},n_{3,1}} (\phi ),\quad  \label{3ca-7}
\end{equation}

\begin{equation}
\left( -\frac{d^{2}}{d\beta^{2}}+\frac{R_{i,n_{2,1},n_{3,1}}-\frac{1}{4}}{\sin ^{2}\beta}%
+\frac{B_{n_{1,1}}-\frac{1}{4}}{ \cos^2\beta } \right) Q_{m,i,\tilde n_1} (\beta )=C_{m,i,\tilde n_1}
 Q_{m,i,\tilde n_1} (\beta ),\qquad  \label{3ca5-8}
\end{equation}%

\begin{equation}
\left( -\frac{d^{2}}{d\theta ^{2}}+\frac{C_{m,i,\tilde n_1}-\frac{1}{4}}{\sin ^{2}\theta }%
+\frac{B_{n_{1,2}}-\frac{1}{4}}{ \cos^2\theta} \right) \Theta_{j,m,i,n_{1,2},\tilde n_1} (\theta )=D_{j,m,i,n_{1,2},\tilde n_1} 
\Theta_{j,m,i,n_{1,2},\tilde n_1} (\theta ),\qquad  \label{3ca-8}
\end{equation}%

\begin{equation}
\left( -\frac{d^{2}}{d\alpha ^{2}}+\frac{D_{j,m,i,n_{1,2},\tilde n_1}-\frac{1}{4}}{\sin ^{2}\alpha }%
\right) G_{\ell,j,m,i,n_{1,2},\tilde n_1} (\alpha )=A_{\ell,j,m,i,n_{1,2},\tilde n_1} 
G_{\ell,j,m,i,n_{1,2},\tilde n_1} (\alpha ),\qquad  \label{3ca-9}
\end{equation}%
and 
\begin{equation}
\left( -\frac{d^{2}}{dr^{2}}+\omega ^{2}r^{2}+\frac{\mu +A_{\ell,j,m,i,n_{1,2},\tilde n_1}-\frac{1}{4}}{r^{2}%
}\right) F_{k,\ell,j,m,i,n_{1,2},\tilde n_1}(r)=E_{k,\ell,j,m,i,n_{1,2},\tilde n_1} \  F_{k,\ell,j,m,i,n_{1,2},\tilde n_1
}(r) \ .  \label{3ca-10}
\end{equation}


The regular solutions of  (\ref{3ca-7})  on the interval $]0,\pi/2[$,  with Dirichlet conditions at the boundaries, are given  by the expressions  \cite{Perelomov1983,BLL2009}
\begin{equation}
 \zeta_{i,n_{2,1},n_{3,1}}(\phi) = (\sin \phi )^{b_{n_{2,1}}+\frac{1}{2}}(\cos \phi )^{
b_{n_{3,1}}+\frac{1}{2}} P_{i}^{(b_{n_{2,1}},b_{n_{3,1}})}(\cos 2\phi ),\qquad 
          \qquad i=0,1,2,...
\label{3ca-22} 
\end{equation}
in terms of the Jacobi Polynomials, and  are  associated to the eigenvalues 
\begin{equation}
R_{i,n_{2,1},n_{3,1}}=r_{i,n_{2,1},n_{3,1}}^2, \qquad r_{i,n_{2,1},n_{3,1}}=(2 i + 1 + b_{n_{2,1}}+b_{n_{3,1}})
  \qquad i=0,1,2,..  \ . \label{3ca-21}
\end{equation}
Here $ b_{n_{2,1}}$ and $ b_{n_{3,1}}$ are  taken from  equations (\ref{3cb-22-bis},\ref{3cb-22-ter}).

The second angular equation for the polar angle $\beta$ reads : 
\begin{equation}
\left( -\frac{d^{2}}{d\beta ^{2}}+\frac{r_{i,n_{2,1},n_{3,1}}^{2}-\frac{1}{4}}{\sin^{2}\beta }+\frac{b_{n_{1,1}}^2-\frac{1}{4}}{ \cos ^2\beta} -C_{m,i,\tilde n_1} \right) Q_{m,i,\tilde n_1} (\beta ) =0 \ .  \label{3ca-24}
\end{equation}%
Its regular solutions    in $]0,\pi/2[$ with Dirichlet conditions are
 \begin{eqnarray}
 Q_{m,i,n_{1,2},\tilde n_1} (\beta )&=&(\sin \beta )^{r_{i,n_{2,1},n_{3,1}}+\frac{1}{2}}(\cos \beta )^{b_{n_{1,1}}+
\frac{1}{2}}P_{m}^{(r_{i,n_{2,1},n_{3,1}},b_{n_{1,1}})}(\cos 2\beta ),\qquad  \label{3ca-17} \\ 
\quad 0 &\leq &\beta \leq \frac{\pi }{2},\quad m=0,1,2,... \ .
\end{eqnarray}
 They are associated to the eigenvalue :
\begin{equation}
C_{m,i,\tilde n_1}=c_{m,i,\tilde n_1}^2, \qquad c_{m,i,\tilde n_1}=(2 m + 1 +  r_{i,n_{2,1},n_{3,1}}+ b_{n_{1,1}}) \ \qquad m=0,1,2,..  \ . 
\end{equation}
 Taking into account Eq.(\ref{3ca-21}) we have  :
\begin{equation} 
c_{m,i,\tilde n_1}=(2 i + 2 m +2  + b_{n_{1,1}}+ b_{n_{2,1}}+b_{n_{3,1}})
  \qquad i=0,1,2,... \quad  m=0,1,2,..  \ . \label{cmna1}
\end{equation} 

The regular solutions Eq.(\ref{3ca-8}) in $]0,\pi/2[$ with Dirichlet conditions read, taking into account Eqs.(\ref{3cbb-22-bis},
\ref{cmna1}),
 \begin{eqnarray}
\Theta _{j,m,i,n_{1,2},\tilde n_1}(\theta ) &=&(\sin \theta )^{c_{m,i,\tilde n_1}+\frac{1}{2}}(\cos \theta )^{b_{n_{1,2}}+\frac{1}{2}}P_{j}^{(c_{m,i,\tilde n_1},b_{n_{1,2}})}(\cos 2\theta ),\qquad  \label{3ga-17} \\
\quad 0 &\leq &\theta \leq \frac{\pi }{2},\quad j=0,1,2,... \ .     \nonumber      
\label{thetapb}
\end{eqnarray}
The eigenvalues  $D_{j,m,i,n_{1,2},\tilde n_1}$ of Eq.(\ref{3ca-8}) are given by 
\begin{equation}
D_{j,m,i,n_{1,2},\tilde n_1}=d_{j,m,i,n_{1,2},\tilde n_1}^2 \ ,  \qquad  d_{j,m,i,n_{1,2},\tilde n_1}=(2 j +c_{m,i,\tilde n_1}+b_{n_{1,2}}+1) , \  j=0,1,2,... \quad ,
\label{Dmna}
\end{equation}
and taking into account (\ref{cmna1})
\begin{eqnarray}
d_{j,m,i,n_{1,2},\tilde n_1}&=&(2 j +2 m +2 i +b_{n_{1,1}}+b_{n_{2,1}} + b_{n_{3,1}}+b_{n_{1,2}}+3) \ , \label{Dmna1} \\
 & &  j=0,1,2,..., \quad m=0,1,2,...,  \quad  i=0,1,2,.. \quad  .\nonumber 
\end{eqnarray}
The regular eigensolutions and corresponding eigenvalues of  Eq.(\ref{3ca-9}) in the interval $]0,\pi[$
read, respectively \cite{BLL2009}, 
\begin{eqnarray}
G _{\ell,j,m,i,n_{1,2},\tilde n_1}(\alpha ) &=&(\sin \alpha )^{d_{j,m,i,n_{1,2},\tilde n_1}+\frac{1}{2}}C_{\ell}^{(d_{j,m,i,n_{1,2},\tilde n_1}+%
\frac{1}{2})}(\cos \alpha ),\qquad \ell=0,1,2,...,\quad  \label{3ca-30} \\
A_{\ell,j,m,i,,n_{1,2},\tilde n_1} &=&a_{\ell,j,m,i,n_{1,2},\tilde n_1}^2, \label{aas} \\
 \quad a_{\ell,j,m,i,n_{1,2},\tilde n_1} & = &
\left( \ell+d_{j,m,n,i,n_{1,2},\tilde n_1}+\frac{1}{2} \right) \quad  \,\, \ell=0,1,2,...\ ,  \nonumber
\end{eqnarray}
and taking into account (\ref{Dmna1})
\begin{eqnarray}
a_{\ell,j,m,i,n_{1,2},\tilde n_1} & = &
\left(\ell+ 2 j +2 m +2 i +\sum_{M=1}^3 b_{n_{M,1}} +b_{n_{1,2}}+\frac{7}{2}\right) , \nonumber\\
 & &  \ell=0,1,2,... , j=0,1,2,... , m=0,1,2,... \ ,i=0,1,2,... \ .
\label{Dmna2}
\end{eqnarray}

Our choices $b_{n_{M,1}} >0$,\ $ M=1,2,3,$ and $ b_{n_{1,2}} >0 $  ensure  that the Hamiltonians of Eqs.(\ref{3ca-7},\ref{3ca5-8},\ref{3ca-8},\ref{3ca-9}) are   self-adjoint operators.

Finally, the reduced radial equation reads 
\begin{equation}
\left( -\frac{d^{2}}{dr^{2}}+\omega ^{2}r^{2}+\frac{\mu +A_{\ell,j,m,i,n_{1,2},\tilde n_1}-\frac{1}{4}}{%
r^{2}}-E_{k,\ell,j,m,i,n_{1,2},\tilde n_1} \right) F_{k,\ell,j,m,i,n_{1,2},\tilde n_1}(r)=0.  \label{3ca-32}
\end{equation}%

 We introduce the auxiliary parameter  $\kappa _{\ell,j,m,i,n_{1,2},\tilde n_1}$ defined by
\begin{equation}
\kappa _{\ell,j,m,i,n_{1,2},\tilde n_1}^{2}=\mu + A_{\ell,j,m,i,n_{1,2},\tilde n_1} \ , \qquad\kappa _{\ell,j,m,i,n_{1,2},\tilde n_1}=\sqrt{\mu + A_{\ell,j,m,i,n_{1,2},\tilde n_1} } \ .
\label{3ca-33}
\end{equation}
 The solution of the radial equation (\ref{3ca-32})  reads  \cite{BLL2009}

\begin{equation}
F_{k,\ell,j,m,i,n_{1,2},\tilde n_1}(r)=r^{\kappa _{\ell,j,m,i,n_{1,2},\tilde n_1}+\frac{1}{2}}\exp \left(-\frac{\omega r^{2}}{2}\right) L_{k}^{(\kappa _{\ell,j,m,i,n_{1,2},\tilde n_1})}(\omega r^{2}),\qquad k=0,1,2... \ ,  \label{3ca37}
\end{equation}
$L_k^{\kappa}$ being  the generalized Laguerre polynomials. The  eigenenergies are given by
\begin{equation}
E_{k,\ell,j,m,i,n_{1,2},\tilde n_1}=2\omega (2k+\kappa _{\ell,j,m,i,n_{1,2},\tilde n_1}+1),\qquad k=0,1,2....  \label{3ca-38}
\end{equation}

Note that the reduced radial equation (\ref{3ca-32})
 is nothing but the usual 3-dimensional harmonic oscillator equation,
 $(\mu + A_{\ell,j,m,i,n_{1,2},\tilde n_1}-1/4)/r^2$ replacing the centrifugal barrier. The 
 square integrable solutions are well known, putting a limit on the coefficient 
 of the $1/r^2$ term, namely  $(\mu + A_{\ell,j,m,i,n_{1,2},\tilde n_1}) >0.$
Note that taking $\mu +A_{\ell,j,m,i,n_{1,2},\tilde n_1} =0$ leads to several self-adjoint extensions differing by a phase.
This fact has been discussed in \cite{BLL2009}. More details can be found in \cite{basu,giri}.
It has to be noted that for attractive centrifugal barriers,  $\mu +A_{\ell,j,m,i,n_{1,2},\tilde n_1} <0$, the problem of collapse appears,  unless regularization procedures are carried out  \cite{case,gupta,camblong,YLL2013}.

Taking into account the definition of $A_{\ell,j,m,i,n_{1,2},\tilde n_1}$,  Eq.(\ref{aas},\ref{Dmna2}), we have
\begin{eqnarray}
& & \mu +A_{\ell,j,m,i,n_{1,2},\tilde n_1} =  \mu +\left( \ell+2 j +2 m + 2 i  + 3 \sum_{M=1}^3 ( n_{M,1} + a_{M,1})
  + 3  n_{1,2} + 3 a_{1,2} + \frac{19}{2} \right)^2 >0 
 \nonumber\\ 
 & & \forall \ell \geq 0,  \forall j\geq 0,\forall m\geq 0, \forall i\geq 0, \forall n_{1,2} \geq 0,  \forall n_{1,1} \geq 0 \ \forall n_{2,1} \geq 0, \forall n_{3,1} \geq 0 \label{3ca-32bispp}
\end{eqnarray}
for every positive $\mu$. 

The quantity $\mu +A_{\ell,j,m,i,n_{1,2},\tilde n_1}$ is minimal for  $\tilde n_1=\tilde {0},n_{1,2}=0,i=0,j=0,m=0,\ell=0$ and $a_{M,1}=0 \ \  (\forall M=1,2,3) , a_{1,2}=0$ ( we recall that $a_{M,1},M=1,2,3,  a_{1,2}\geq 0$
see (\ref{3cb-22-ter},\ref{3cbb-22-ter})). The positivity of $\mu+A_{\ell,j,m,i,n_{1,2},\tilde n_1}$ puts constraints on negative values of  $\mu$,  namely 
\begin{equation}
 -  \left(\frac{19}{2}\right)^2 <  \mu \leq  0  \ .
\end{equation}

Collecting all pieces, we conclude that  the physically acceptable (non  normalized) solutions of 
the Schr\"odinger equation  (\ref{3cb-49}) are  given, in a compact and symmetrized form, by  
\begin{eqnarray}
& &\Psi_{k,\ell,j,m,i,n_{1,2},\tilde n_1}(r,\alpha,\theta,\beta,\phi,\varphi_{1,2}, \tilde \varphi_1) = 
 r^{\sqrt{\mu+ (19/2 + 3 a_{1,2} + 3 n_{1,2} + \sum_{M=1}^3 (3 a_{M,1} + 3 n_{M,1})  + \ell + 2 j + 2 m + 2 i )^2 \ }-7/2 } \nonumber\\
 & & \times 
L_k^{\sqrt{\mu+ (19/2 + 3 a_{1,2} + 3 n_{1,2} + \sum_{M=1}^3 (3 a_{M,1} + 3 n_{M,1})  + \ell + 2 j + 2 m  + 2 i )^2}}
(\omega r^{2}) \ \exp \left(-\frac{\omega r^{2}}{2}\right)  \nonumber\\
& & \times (\sin \alpha )^{ 6 + 3 a_{1,2} + 3 n_{1,2} + \sum_{M=1}^3 (3 a_{M,1} + 3 n_{M,1})  + 2 i +2  j + 2 m 
} \nonumber\\
 & &  \times C_{\ell}^{19/2 +3 a_{1,2} + 3 n_{1,2} + \sum_{M=1}^3 (3 a_{M,1} + 3 n_{M,1})  + 2 i +2  j + 2 m
}(\cos \alpha ) \nonumber\\
& & \times (\sin \theta )^{9/2+\sum_{M=1}^3 (3 a_{M,1} + 3 n_{M,1})  +  2 i + 2 m} (\cos \theta )^{ 3/2 + 3 a_{1,2} + 3 n_{1,2}} \nonumber\\
 & &  \times P_{j}^{ 13/2 + \sum_{M=1}^3 (3 a_{M,1} + 3 n_{M,1})  +  2 i + 2 m, 
    3/2  + 3 a_{1,2} + 3n_{1,2}}(\cos 2\theta ) \nonumber\\
 & &\times  (\sin \beta )^{3 +2 i+ 3 a_{2,1} + 3 a_{3,1} + 3 n_{2,1} + 3 n_{3,1}}(\cos \beta )^{ 3/2+ 3 a_{1,1} + 3 n_{1,1}}P_{m}^{4 +3 a_{2,1} + 3 a_{3,1} + 3 n_{2,1} + 3 n_{3,1}, 3/2 +3 a_{1,1} + 3 n_{1,1}}(\cos 2\beta ) \nonumber\\ 
& &  \times (\sin \phi )^{3 a_{2,1} + 3 n_{2,1} + 3/2}(\cos \phi )^{3 a_{3,1}+ 3 n_{3,1} + 3/2} 
  P_{i}^{3/2 + 3 \ a_{2,1} + 3 \ n_{2,1}, 3/2 + 3 \ a_{3,1} + 3 n_{3,1}}(\cos 2\phi ) \nonumber\\
& & \times \vert \sin 3\varphi_{1,2} \vert^{\frac{1}{2}+ a_{1,2}} C_{n_{1,2}}^{(\frac{1}{2}+
a_{1,2})}(\cos 3 \varphi_{1,2} ) 
\prod_{M=1}^3  \vert \sin 3\varphi_{M,1} \vert ^{\frac{1}{2}+ a_{M,1}}C_{n_{M,1}}^{(\frac{1}{2}+
a_{M,1})}(\cos 3\varphi_{M,1} )   \ ,   \label{compb} 
\end{eqnarray}
 with
\begin{eqnarray}
k &=&0,1,2,..., \ell=0,1,2,..., j=0,1,2,.., m=0,1,2,.., i=0,1,2,..., \nonumber\\
n_{1,2} &=&0,1,2,..., n_{1,1}=0,1,2,...,   n_{2,1}=0,1,2,..., n_{3,1}=0,1,2,...   \nonumber \\
 &&  (\forall M) \quad  (1 \leq M \leq 3) \ \quad 0 \leq \varphi_{M,1} \leq \frac{\pi }{3},\quad 0 \leq \varphi_{1,2} \leq \frac{\pi }{3}
     \quad 0\leq \phi \leq \frac{\pi }{3}, \nonumber\\
     & &   \quad 0 \leq \beta \leq \frac{\pi }{2},  \quad 0 \leq \theta \leq \frac{\pi }{2},  \quad 0 \leq \alpha \leq \pi
\quad   0 \leq r \leq  \infty      \nonumber\\
& &(\forall M) \quad  (1 \leq M \leq 3) \quad  a_{M,1}=\frac{1}{2}\sqrt{1+2\lambda_{M,1} } \ ,\quad   \quad a_{1,2}=\frac{1}{2}\sqrt{
1+2\lambda_{1,2}} \ . \nonumber    
\end{eqnarray}
 It has  to be noticed   that, for  Bose  statistics,  a $\delta$ pathology  occurs in (\ref{compb})  for 
 $a_{M,1}=1/2$ ($\lambda_{M,1}=0, M=1,2,3$) and $a_{1,2}=1/2$ ($\lambda_{1,2}=0$).

The normalization constant $N_{k,\ell,j,m,i,n_{1,2},\tilde n_1}$ of the wave function $ \Psi_{k,\ell,j,m,i,n_{1,2},\tilde n_1}$, Eq.(\ref{compb}), can be  calculated from
\begin{eqnarray}
&& \int_{0}^{+\infty }r^{8}\ dr \int_{0}^{\pi }\sin^7 \alpha \ d\alpha 
\int_{0}^{\pi/2} \sin^5 \theta \cos \theta  \ d\theta \int_{0}^{\pi/2} \sin^3 \beta \cos \beta \ d\beta
\int_{0}^{\pi/2} \sin 2 \phi  \   \ d\phi  \int_0^{\pi/3}d\varphi_{1,2} \nonumber\\
& &  \prod_{M=1}^3   \int_0^{\frac{\pi }{3}}d\varphi_{M,1}
\Psi _{k,\ell,j,m,i,n_{1,2},\tilde n_1}(r,\alpha,\theta ,\beta,\phi,\varphi_{1,2},\tilde \varphi_1 )\Psi_{k^{\prime },\ell^{\prime },j^{\prime },m^{\prime },i^{\prime },n_{1,2}^{\prime },\tilde n_1^{\prime }}
(r,\alpha,\theta ,\beta,\phi,\varphi_{1,2},\tilde \varphi_1)  \nonumber\\
& =& \delta_{k,k^{\prime }}\delta _{\ell,\ell^{\prime }}\delta _{j,j^{\prime }}\delta _{m,m^{\prime }} 
  \delta _{i,i^{\prime }}   \delta _{n_{1,2},n_{1,2}^{\prime }} \  \delta _{n_{1,1},n_{1,1}^{\prime }} \ 
\delta _{n_{2,1},n_{2,1}^{\prime }} \ \delta _{n_{3,1},n_{3,1}^{\prime }} \ N_{k,\ell,j,m,i,n_{1,2},\tilde n_1} .
\end{eqnarray}%
Use is made  of the orthogonality properties of  Gegenbauer, Jacobi and Laguerre polynomials \cite{Abramowitzbook}.  
The normalization constant  $N_{k,\ell,j,m,i,n_{1,2},\tilde n_1}$ can be worked out analytically. Its expression involves products of norms of  Gegenbauer, Jacobi and  Laguerre polynomials.  The eigenenergies are :

\begin{equation}
\frac{E_{k,\ell,j,m,i,n_{1,2},\tilde n_1}}{2\omega} =2k+1+\sqrt{\mu+\left(\ell+2j+2m+2i+3n_{1,2}+3a_{1,2}+
3 \sum_{M=1}^3 (n_{M,1}+a_{M,1})+\frac{19}{2}\right)^2 }    \label{ener9} 
\end{equation}
\begin{eqnarray}
& &  k=0,1,2.... ,  \ell=0,1,2,.... , j=0,1,2,.... , m=0,1,2,.... , i=0,1,2,.... , \nonumber\\
& & n_{1,2}=0,1,2,.... , n_{1,1}=0,1,2,.... , n_{2,1}=0,1,2,.... ,,n_{3,1}=0,1,2,.... \ ,\nonumber\\
& & (\forall M) \quad  (1 \leq M \leq 3) \quad  a_{M,1}=\frac{1}{2}\sqrt{1+2\lambda_{M,1} } \ ,\quad   \quad a_{1,2}=\frac{1}{2}\sqrt{1+2\lambda_{1,2}} \nonumber . 
\end{eqnarray}
   
Setting $\mu=0$ we have,
\begin{eqnarray}
E_{k,\ell,j,m,i,n_{1,2},\tilde n_1} = 2\omega \left\{\frac{21}{2} +3n_{1,2}+3a_{1,2} 
 +3 \sum_{M=1}^3 (n_{M,1}+a_{M,1})+ \ell +  2k +2j+2m+2i \right \} \ .
\label{enerp9}
\end{eqnarray}

 This  expression is equivalent to  Eq.(\ref{enercartp}). Indeed, the energy spectra are similar 
 if we identify $M$ (of Eq.\ref{enercartp})) and $\ell$ (of Eq.(\ref{enerp9})). Then, setting 
 $L=2k+2j+2m+2i$ for Eq.(\ref{enerp9}) and $L=2  k_{n_{1,2}} + 2 \sum_{i=1}^3  k_{n_{i,1}} $ for Eq.(\ref{enercartp}) we obtain identical expressions of the energies, Eqs.(\ref{enercartp},\ref{enerp9}).
 The energy spectra are thus the same.

\section{The general case of $N=3^k$ particles  ( $ k \geq 2$) }

The same procedure can be generalized to  $N=3^k, \  k \geq 2. $ As before, the $N$ particles, with coordinates  $x_i,i=1,2,\dots,3^k$ are confined in a harmonic well.  Then, the  $N$ particles are clustered in $3^{k-1}$ clusters of 3 particles (clusters of the first level).  The $\ell^{\rm th}, (\ell=1,,2,\dots,3^{k-1})$ cluster of the first level incorporates
the three particles $x_i,i=3\ell-2,3\ell-1,3\ell$.
At the next step, the "first level clusters" are clustered in "second level clusters" of 3 clusters. In each of these 
 "second level clusters" the "first level clusters" interact via a 2-body Calogero-type of potential given in terms of their centre of mass coordinates,labeled $w_{i,1},i=1,2,\ldots,3^{k-2}$.
  The $\ell^{\rm th}, (\ell=1,2,\dots,3^{k-2})$ cluster of the second  level incorporates
the three clusters of first level  positioned at  $w_{i,1},i=3\ell-2,3\ell-1,3\ell$.
  This procedure is generalized by clustering further  "second level clusters" for $k \geq 3$, etc...
 Finally, a non-translationally invariant $N$-body potential is added, with coupling constant $\mu$,
  namely $\mu/(\sum_{\ell=1}^{3^k} x_{\ell}^2)$.
This quantum $N$-body system is represented by the following  Hamiltonian :

\begin{eqnarray}
H &= &\sum_{\ell=1}^{3^k}\left( -\frac{\partial ^{2}}{\partial x_{\ell}^{2}}%
+\omega ^{2}  x_{\ell}^{2}\right) + \frac{\mu}{ \sum_{\ell=1}^{3^k} x_{\ell}^2}  \nonumber\\
 & + & \sum_{\ell=1}^{3^{k-1}} \lambda_{\ell,1} \sum_{3 \ell-2 \leq i < j \leq 3 \ell } \frac{1}{(x_{i}-x_{j})^{2}} 
 +  \sum_{\ell=1}^{3^{k-2}} \lambda_{\ell,2} \sum_{3 \ell-2 \leq i < j \leq 3 \ell } \frac{1}{(w_{i,1}-w_{j,1})^{2}} \nonumber\\
&+ & \sum_{\ell=1}^{3^{k-3}}  \lambda_{\ell,3} \sum_{3 \ell-2 \leq i < j \leq 3 \ell }
 \frac{1}{(w_{i,2}-w_{j,2})^{2}  }  
 +  ......... +     \sum_{1 \leq i < j \leq 3} \frac{\lambda_{1,k}}{(w_{i,k-1}-w_{j,k-1})^{2}  }  \ ,
\label{3cb-1}
\end{eqnarray}
where we have defined the centres  of mass of three-body clusters, clusters of "type 1" : 
\begin{equation}
w_{\ell,1} =  \frac{x_{3 \ell-2}+x_{3 \ell-1} + x_{3 \ell}  }{\sqrt{3}} \qquad\quad \ell=1,2,3,...,3^{k-1}  \ .
\end{equation}
The $x_{\ell}$'s are the coordinates of the particles $(\ell=1,2,..,3^k)$.
The  centres  of mass of nine-body clusters,    clusters of "type 2", read 
\begin{equation} 
w_{\ell,2}  =  \frac{w_{3 \ell-2,1}+w_{3 \ell-1,1}+ w_{3 \ell,1}  }{\sqrt{3}} \qquad\quad \ell=1,2,3,...,3^{k-2}  \ ,
\end{equation}
etc...,  and the  centres  of mass of $3^n$-body clusters, clusters of "type n",display : 
 \begin{equation} 
w_{\ell,n}  =  \frac{w_{3 \ell-2,n-1}+w_{3 \ell-1,n-1} + w_{3 \ell,n-1} }{\sqrt{3}} \qquad\quad \ell=1,2,3,...,3^{k-n}, \quad n \geq 2 \ .
\label{coord}
\end{equation}

Starting from the Hamiltonian, Eq.(\ref{3cb-1}), we introduce the change of coordinates
\begin{eqnarray}
 u_{\ell,1}&=&\frac{1}{\sqrt{2}}(x_{3 \ell -2}-x_{3 \ell -1}),   v_{\ell,1} = \frac{1}{\sqrt{6}}(x_{3 \ell -2}+x_{3 \ell -1}-2x_{3 \ell}) 
 \label{3cb-16ter} \\
& &   w_{\ell,1}= \frac{1}{\sqrt{3}} (x_{3 \ell-2}+x_{3 \ell-1} + x_{3 \ell}) \ , \qquad\quad
 \ell=1,2,...,3^{k-1} \nonumber .
\end{eqnarray}

The transformed Hamiltonian reads:
\begin{eqnarray}
H &= &\sum_{\ell=1}^{3^{k-1}} \left[ -\frac{\partial ^{2}}{\partial u_{\ell,1}^2}
-\frac{\partial^{2}}{\partial v_{\ell,1}^2}  -\frac{\partial ^{2}}{\partial w_{\ell,1}^{2}}+
 \omega^2 [ u_{\ell,1}^2+v_{\ell,1}^2+w_{\ell,1}^2]
+ \frac{9\lambda_{\ell,1} [u_{\ell,1}^2+v_{\ell,1}^2]^2}{2\left[u_{\ell,1}^3 -   3 u_{\ell,1} v_{\ell,1}^2 \right]^{2}}  \right]  \nonumber\\
  & +&  \sum_{\ell=1}^{3^{k-2}} \lambda_{\ell,2}
 \sum_{3 \ell-2 \leq i < j \leq 3 \ell } \frac{1}{(w_{i,1}-w_{j,1})^{2}} 
 +  \sum_{\ell=1}^{3^{k-3}} \lambda_{\ell,3} \sum_{3 \ell-2 \leq i < j \leq 3 \ell }  \frac{1}{(w_{i,2}-w_{j,2})^{2}}  
  \nonumber\\
 & + & .........  +  \sum_{1 \leq i < j \leq 3  } \frac{ \lambda_{1,k}}{(w_{i,k-1}-w_{j,k-1})^{2}  } 
+ \frac{\mu}{ \sum_{\ell=1}^{3^{k-1}} [u_{\ell,1}^2+v_{\ell,1}^2+w_{\ell,1}^2]}   \ .
\label{3cb-481}
\end{eqnarray}

Then we introduce the transformation

\begin{eqnarray}
 u_{\ell,2}&=&\frac{1}{\sqrt{2}}(w_{3 \ell -2,1}-w_{3 \ell -1,1}),   v_{\ell,2} = \frac{1}{\sqrt{6}}(w_{3 \ell -2,1}+
 w_{3 \ell -1,1}-2w_{3 \ell,1}) 
 \label{3cb-16terr} \\
& &   w_{\ell,2}= \frac{1}{\sqrt{3}} (w_{3 \ell-2,1}+w_{3 \ell-1,1} + w_{3 \ell,1}) \ , \qquad\quad
 \ell=1,2,...,3^{k-2} \  , \nonumber
\end{eqnarray}

and obtain the Hamiltonian

\begin{eqnarray}
H &= &\sum_{\ell=1}^{3^{k-1}} \left[ -\frac{\partial ^{2}}{\partial u_{\ell,1}^2}
-\frac{\partial^{2}}{\partial v_{\ell,1}^2} + \omega^2 [ u_{\ell,1}^2+v_{\ell,1}^2] +
  \frac{9\lambda_{\ell,1} [u_{\ell,1}^2+v_{\ell,1}^2]^2}{2\left[u_{\ell,1}^3 -   3 u_{\ell,1} v_{\ell,1}^2 \right]^{2}} \right] \nonumber\\
  &+& \sum_{\ell=1}^{3^{k-2}}  \left[-\frac{\partial ^{2}}{\partial u_{\ell,2}^2}-\frac{\partial^2}{\partial v_{\ell,2}^2} 
-\frac{\partial^2}{\partial w_{\ell,2}^2}+ \omega^2 [ u_{\ell,2}^2+v_{\ell,2}^2+w_{\ell,2}^2] 
 +  \frac{9\lambda_{\ell,2} [u_{\ell,2}^2+v_{\ell,2}^2]^2}{2\left[u_{\ell,2}^3 -   3 u_{\ell,2} v_{\ell,2}^2 \right]^{2}}  \right]  \nonumber\\  
 &  + &   \sum_{\ell=1}^{3^{k-3}} \lambda_{\ell,3} \sum_{3 \ell-2 \leq i < j \leq 3 \ell }
 \frac{1}{(w_{i,2}-w_{j,2})^{2}} 
 + \sum_{\ell=1}^{3^{k-4}}  \lambda_{\ell,4}  \sum_{3 \ell-2 \leq i < j \leq 3 \ell } \frac{1}{(w_{i,3}-w_{j,3})^{2}  } 
  +  ......... \nonumber\\
 &+ &    \sum_{1 \leq i < j \leq 3 } \frac{  \lambda_{1,k}}{(w_{i,k-1}-w_{j,k-1})^{2}  } 
 + \frac{\mu}{\sum_{\ell=1}^{3^{k-1}} [ u_{\ell,1}^2+v_{\ell,1}^2]+ \sum_{\ell=1}^{3^{k-2}} [u_{\ell,2}^2+v_{\ell,2}^2+w_{\ell,2}^2]}  \ . \nonumber
\end{eqnarray}
 The procedure of successive coordinate transformations is repeated and, at the $n^{{\rm th}}$ step, $n=2,3,..,k$, we have
\begin{eqnarray}
 u_{\ell,n}&=&\frac{1}{\sqrt{2}}(w_{3 \ell -2,n-1}-w_{3 \ell -1,n-1}),   v_{\ell,n} = \frac{1}{\sqrt{6}}(w_{3 \ell -2,n-1}+w_{3 \ell -1,n-1}- 2w_{3 \ell,n-1})  \nonumber\\
w_{\ell,n}&= &    \frac{1}{\sqrt{3}} (w_{3 \ell-2,n-1}+w_{3 \ell-1,n-1} + w_{3 \ell,n-1}), \ 
 \ell=1,2,...,3^{k-n}, \  n=2,3,...,k \ , \nonumber
\end{eqnarray}

 and the Hamiltonian reads as : 
\begin{eqnarray}
H &= &\sum_{m=1}^{n} \left \{ \sum_{\ell=1}^{3^{k-m}} \left[ -\frac{\partial ^{2}}{\partial u_{\ell,m}^2}
-\frac{\partial^{2}}{\partial v_{\ell,m}^2} + \omega^2 [ u_{\ell,m}^2+v_{\ell,m}^2] +
  \frac{9\lambda_{\ell,m} [u_{\ell,m}^2+v_{\ell,m}^2]^2}{2\left[u_{\ell,m}^3 -   3 u_{\ell,m} v_{\ell,m}^2 \right]^{2}} \right]   \right \}  \nonumber\\
   & + & \sum_{\ell=1}^{3^{k-n}}\left(- \frac{\partial^2}{\partial w_{\ell,n}^2}+ \omega^2  w_{\ell,n}^2 \right)
  + (1-\delta_{n,k}) \  \sum_{m=n+1}^{k} \left\{    \sum_{\ell=1}^{3^{k-m}}
  \lambda_{\ell,m} \sum_{3 \ell-2 \leq i < j \leq 3 \ell } \frac{1}{(w_{i,m-1}-w_{j,m-1})^{2}} \right\} \nonumber\\
& +& \frac{\mu}{(\sum_{m=1}^n \sum_{\ell=1}^{3^{k-m}} [ u_{\ell,m}^2+v_{\ell,m}^2])+\sum_{\ell=1}^{3^{k-n}} w_{\ell,n}^2}  \  
\label{3cb-483}
\end{eqnarray}
where  $\delta_{n,k}$ denotes the Kronecker symbol. The number of partial derivatives is equal to $3^k$ since 
\begin{eqnarray}
& & 2 \  (3^{k-1} + 3^{k-2} + ... + 3^{k-n}) + 3^{k-n}= 2 \ 3^{k-n} (1 + 3 + 3^2 + ... + 3^{n-1}) + 3^{k-n} \nonumber\\
& & = 2 \ \frac{3^{k-n} (3^n-1)}{2} + 3^{k-n} = 3^k \ .
\end{eqnarray}

At the end we have, for $n=k$  : 
\begin{eqnarray}
H &= &\sum_{m=1}^{k}  \sum_{\ell=1}^{3^{k-m}} \left[ - \frac{\partial ^{2}}{\partial u_{\ell,m}^2}
-\frac{\partial^{2}}{\partial v_{\ell,m}^2}+\omega^2 [ u_{\ell,m}^2+v_{\ell,m}^2]+
 \frac{9\lambda_{\ell,m} [u_{\ell,m}^2+v_{\ell,m}^2]^2}{2\left[u_{\ell,m}^3 -   3 u_{\ell,m} v_{\ell,m}^2 \right]^{2}}  \right] \nonumber\\
 &-&\frac{\partial ^{2}}{\partial w_{1,k}^2}+ \omega^2   w_{1,k}^2 
+ \frac{\mu}{(\sum_{m=1}^k \sum_{\ell=1}^{3^{k-m}}  [u_{\ell,m}^2+v_{\ell,m}^2])+w_{1,k}^2} \ .
\label{3cb-4833}
\end{eqnarray}
Setting
\begin{eqnarray}
u_{\ell,m} & =& r_{\ell,m} \sin \varphi_{\ell,m} \quad v_{\ell,m}= r_{\ell,m} \cos \varphi_{\ell,m}, \quad  \ell=1,2,... ,3^{k-m}, \quad m=1,2,...,k \ , \nonumber\\
0 & \leq  &  r_{\ell,m} < \infty  \ , \quad 0 \leq \varphi_{\ell,m} \leq 2 \pi  \ , \nonumber
\end{eqnarray}

we obtain
\begin{eqnarray}
H &= &\sum_{m=1}^{k}  \sum_{\ell=1}^{3^{k-m}} \left[ - \frac{\partial ^{2}}{\partial r_{\ell,m}^2}
-\frac{1}{r_{\ell,m}} \frac{\partial}{\partial r_{\ell,m}}+\omega^2 r_{\ell,m}^2+ \frac{1}{r_{\ell,m}^2}
\left(-\frac{\partial^2}{\partial \varphi_{\ell,m}^2} +\frac{9\lambda_{\ell,m} }{2 \sin^2 (3 \varphi_{\ell,m})} \right) \right] \nonumber\\
 &-& \frac{\partial ^{2}}{\partial w_{1,k}^2}+ \omega^2  \  w_{1,k}^2 
+ \frac{\mu}{(\sum_{m=1}^k \sum_{\ell=1}^{3^{k-m}}  r_{\ell,m}^2 )+w_{1,k}^2} \ .
\label{3cb-4833p}
\end{eqnarray}

Let us introduce the notation for the set of $(3^k-1)/2 $ variables
\begin{equation}
V_{k}({\bf y}) \equiv \{y_{3^{k-1},1},y_{3^{k-1}-1,1},...,y_{2,1},y_{1,1}, y_{3^{k-2},2},...,y_{1,2},...,
y_{3,k-1},y_{2,k-1},y_{1,k-1},y_{1,k}\}
\end{equation}
For further convenience we introduce the truncated sets:
\begin{equation}
\hat{ V}_{k}({\bf y}) \equiv \{y_{3^{k-1}-1,1},y_{3^{k-1}-2,1},...,y_{2,1},y_{1,1}, y_{3^{k-2},2},...,y_{1,2},...,
y_{3,k-1},y_{2,k-1},y_{1,k-1},y_{1,k}\}
\end{equation}
\begin{eqnarray}
W_{\ell,k-m}({\bf y}) & \equiv &  \{y_{3^{k-1},1},y_{3^{k-1}-1,1},...,y_{2,1},y_{1,1}, y_{3^{k-2},2},...,y_{1,2},...,
 \nonumber\\
 & &  y_{3^m,k-m},y_{3^{m}-1,k-m},....,y_{\ell+1,k-m},y_{\ell,k-m}\}
\end{eqnarray}
and
\begin{eqnarray}
\hat{ W}_{\ell,k-m}({\bf y})  & \equiv  & \{y_{3^{k-1}-1,1},y_{3^{k-1}-2,1},...,y_{2,1},y_{1,1}, y_{3^{k-2},2},...,y_{1,2},...,   \nonumber\\
& &  y_{3^m,k-m},y_{3^{m}-1,k-m},....,y_{\ell+1,k-m},y_{\ell,k-m}\} 
\end{eqnarray}

We have $W_{1,k}({\bf y})  \equiv V_{k}({\bf y})$ and $\hat{W}_{1,k}({\bf y})  \equiv \hat{V}_{k}({\bf y})$, these latter  both sets being defined for $k \ne 1$.

The equation (\ref{3cb-4833p}) suggests that the wave function $\Psi$, solution of $H \Psi=E \Psi$,  can be  factorized as follows
\begin{equation}
\Psi(w_{1,k},V_{k}({\bf r}),V_{k}({\bf \varphi})) =\frac{1}{\sqrt{\prod_{m=1}^k \ \prod_{\ell=1}^{3^{k-m}} r_{\ell,m} \ }}
\times  \chi(w_{1,k},V_{k}({\bf r}))  \times \prod_{m=1}^k \prod_{\ell=1}^{3^{k-m}}
 \Phi_{(\ell,m)}(\varphi_{\ell,m} )  \ .
\label{genpsi}
\end{equation}
This factorization permits to separate the "angular" equations from the "radial" ones.
The $(3^k-1)/2$ angular equations are solved independently: 

\begin{eqnarray}
& & \left( -\frac{d^{2}}{d\varphi_{\ell,m}^{2}}+\frac{9\lambda_{\ell,m} }{2\sin^{2}(3\varphi_{\ell,m}) }%
\right) \Phi_{n_{\ell,m}} (\varphi_{\ell,m} )=B_{n_{\ell,m}} \Phi_{n_{\ell,m}} (\varphi_{\ell,m} ), \qquad\quad   \label{3cb1-7b} \\
& & m=1,2,..,k \qquad\quad  \ell=1,2,..,3^{k-m} \ , \nonumber
\end{eqnarray}
on the interval $]0,\pi/3[$,  with Dirichlet conditions at the boundaries $\varphi_{\ell,m}=0,\pi/3$. The $B_{n_{\ell,m}}$  are the quantized  eigenvalues of  the equations (\ref{3cb1-7b})
respectively  given by 
\begin{eqnarray}
B_{n_{\ell,m}} &= &b_{n_{\ell,m}}^2, \qquad b_{n_{\ell,m}}=3\left( n_{\ell,m} + \frac{1}{2} + a_{\ell,m} \right),\label{3cb-22-bisp} \\
  a_{\ell,m} &  = & \frac{1}{2} \sqrt{1+2\lambda_{\ell,m} }   \ ,  n_{\ell,m}=0,1,2,.... , \  m=1,..,k, \   \ell=1,2,... , 3^{k-m} \ . \label{3cb-22-terp}
\end{eqnarray}

The associated eigensolutions are given in terms of the Gegenbauer polynomials $C_n^{(q)}$ 
\begin{equation}
\Phi_{n_{\ell,m}}(\varphi_{\ell,m} ) = (\sin 3\varphi_{\ell,m} )^{\frac{1}{2}+ a_{\ell,m}} C_{n_{\ell,m}}^{(\frac{1}{2}+
a_{\ell,m})}(\cos 3\varphi_{\ell,m} ),     \ 0\leq \varphi_{\ell,m} \leq \frac{\pi }{3},    
          \ n_{\ell,m}=0,1,2,.. \ .
\label{3cb-222} 
\end{equation}

We have now to solve the following Schr\"odinger equation 
\begin{eqnarray}
& & \left\{ \sum_{m=1}^{k}  \sum_{\ell=1}^{3^{k-m}} \left[ - \frac{\partial ^{2}}{\partial r_{\ell,m}^2}
+\omega^2 r_{\ell,m}^2+ \frac{B_{n_{\ell,m}}-\frac{1}{4}}{r_{\ell,m}^2}- \frac{\partial ^{2}}{\partial w_{1,k}^2}+ \omega^2  \  w_{1,k}^2 \right]\right. \nonumber\\
 && \left. + \frac{\mu}{(\sum_{m=1}^k \sum_{\ell=1}^{3^{k-m}}  r_{\ell,m}^2 )+w_{1,k}^2} -E_{V_k({\bf n})} \right \} \chi_{V_k({\bf n})}(w_{1,k},V_{k}({\bf r})) = 0  \ .
\label{3cb-485}
\end{eqnarray}

The general solution of the  latter equation reads, for $\mu=0$,
\begin{eqnarray}
& &  \chi_{n_{w_{1,k}},V_k(\Lambda),V_k(n)}(w_{1,k},V_{k}({\bf r}))= H_{n_{w_{1,k}}}(\sqrt{\omega}  w_{1,k})  \exp(-\omega w_{1,k}^2/2)  \nonumber\\
&\times  & \prod_{m=1}^k \prod_{\ell=1}^{3^{k-m}} \   r_{\ell,m}^{ b_{n_{\ell,m}}+1/2} L_{\Lambda_{\ell,m}}^{(b_{n_{\ell,m}})}(\omega r_{\ell,m}^2) \exp(-\omega  r_{\ell,m}^{2}/2)  \nonumber\\
 & &   n_{w_{1,k}}=0,1,2,.. \ , \quad \Lambda_{\ell,m}=0,1,2,... \  , \quad  \ell=1,2,..,3^{k-m} \quad \nonumber m=1,2,..,k
\end{eqnarray}
in terms of the Hermite $(H_{n_{w_{1,k}}})$ and the Laguerre polynomials $L_{\Lambda_{\ell,m}}^{(b_{n_{\ell,m}})}$.  
The eigenenergy reads 

\begin{equation}
E_{n_{w_{1,k}},V_k({\bf \Lambda}),V_k({\bf n})}= 2 \omega \left\{ \frac{1}{2} +  n_{w_{1,k}} + \sum_{m=1}^{k}  \sum_{\ell=1}^{3^{k-m}}
\left[1+ 2  \Lambda_{\ell,m} +  3\left( n_{\ell,m} + \frac{1}{2} + a_{\ell,m} \right) \right]\right\} \ .
\label{ener}
\end{equation}
If all the   $a_{\ell,m}$'s, Eq.(\ref{3cb-22-terp}), are equal to, say, $a$, the summation in Eq.(\ref{ener})
leads to a term  $a (N-1)/2$ with $N=3^k$, contributing to the energy Eq.(\ref{ener}). 
If $ (\forall \ell)$ the coupling constant satisfies
$ \lambda_{\ell,m}=\lambda_{1,m}$ and $\lambda_{1,m}=9^{m-1}  \lambda_{1,1}$,  the summation 
 in Eq.(\ref{ener}), for high values of  $\lambda_{1,1}$, leads to 
a number of terms proportional  to  $ a_{1,1}$ equal to  $  3  ( 3^{k-1} + 3^{k} + 3^{k+1}+ 3^{2 k-2}) \simeq  \ 3^k (3^k-1)/2 = N (N-1)/2 $. Therefore we have a term like   $a N  (N-1)/2$  contributing to the energy.

For  $\mu \ne 0$ the  Hamiltonian is not  separable in  $\{w_{1,k},V_k({\bf r})\}$ variables.
As  before, we  introduce   hyperspherical coordinates, taking into account the fact that all components of $V_k({\bf r})$ have positive values : 

\begin{eqnarray}
w_{1,k} &=& r \cos \alpha, \qquad\quad 0  \leq   r <\infty  \ ,    \quad  0 \leq \alpha \leq \pi  \nonumber\\
r_{1,k} &=& r \sin \alpha \cos \beta_{1,k} \ ,  \quad  0 \leq \beta_{1,k} \leq \frac{\pi}{2}, \quad k \ne 1 \nonumber\\
r_{1,k-1} &=& r \sin \alpha \sin \beta_{1,k} \cos   \beta_{1,k-1} \ , \quad  0 \leq \beta_{1,k-1} \leq \frac{\pi}{2}  \nonumber\\
r_{2,k-1} &=& r \sin \alpha \sin \beta_{1,k} \sin   \beta_{1,k-1}  \cos \beta_{2,k-1} \ , \quad  0 \leq \beta_{2,k-1} \leq \frac{\pi}{2}  \nonumber\\
r_{3,k-1} &=& r \sin \alpha \sin \beta_{1,k} \sin   \beta_{1,k-1}  \sin \beta_{2,k-1} \cos \beta_{3,k-1} \ , \quad  0 \leq \beta_{3,k-1} \leq \frac{\pi}{2}  \nonumber\\
 ... & & ... \nonumber\\
r_{1,k-m} & = &  \left[ r \sin \alpha  \prod_{j=1}^m \prod_{i=1}^{3^{j-1}} \sin \beta_{i,k-j+1} \right] \cos \beta_{1,k-m}
\ ,  \quad  0 \leq \beta_{1,k-m} \leq \frac{\pi}{2} \ ,  1 \leq m \leq k-2 \nonumber\\
... &  & ...\nonumber\\
r_{\ell,k-m} & = &   \left[ r \sin \alpha  \prod_{j=1}^m \prod_{i=1}^{3^{j-1}} \sin \beta_{i,k-j+1}  \right] \times  \left( \prod_{j=1}^{\ell-1} \sin  \beta_{j,k-m} \right) \  \cos  \beta_{\ell,k-m} \ \nonumber\\
  & & \quad 0 \leq \beta_{\ell,k-m} \leq \frac{\pi}{2} \ ,  \quad 1 \leq \ell \leq 3^m \ ,, \nonumber\\
... & & ... \nonumber\\
r_{\ell,1} & = & \left[  r \sin \alpha  \prod_{j=1}^{k-1} \prod_{i=1}^{3^{j-1}} \sin \beta_{i,k-j+1} \times  \prod_{j=1}^{\ell-1} 
\sin  \beta_{j,1}\right] \cos  \beta_{\ell,1} \ , \quad  0 \leq \beta_{\ell,1} \leq \frac{\pi}{2} \ , \ell \geq 1   \nonumber\\
... & & ... \nonumber\\
r_{3^{k-1}-1,1} & = & \left[ r \sin \alpha  \prod_{j=1}^{k-1} \prod_{i=1}^{3^{j-1}} \sin \beta_{i,k-j+1} \times  \prod_{j=1}^{3^{k-1}-2} \sin  \beta_{j,1} \right] \cos \beta_{3^{k-1}-1,1} , \quad  0 \leq \beta_{3^{k-1}-1,1} \leq \frac{\pi}{2} \nonumber\\
r_{3^{k-1},1} & = & \left[ r \sin \alpha  \prod_{j=1}^{k-1} \prod_{i=1}^{3^{j-1}} \sin \beta_{i,k-j+1} \times  \prod_{j=1}^{3^{k-1}-2} \sin  \beta_{j,1} \right] \sin \beta_{3^{k-1}-1,1}
\label{3ca-18biss}
\end{eqnarray}

The Schr\"{o}dinger equation (\ref{3cb-485}) is then written as :
\begin{eqnarray}
&&\left\{ -\frac{\partial ^{2}}{\partial r^{2}}-\frac{3^k-1}{2 r}\frac{\partial }{%
\partial r}+\omega ^{2}r^{2}+\frac{\mu }{r^{2}}  +  \frac{1}{r^{2}}\left[
 -\frac{\partial ^{2}}{\partial \alpha ^{2}}   -\frac{3^k-3}{2} \cot  \alpha \frac{\partial }{\partial \alpha } \right.\right. \nonumber\\
 & & \left.\left.  +\frac{1}{\sin ^{2} \alpha }  \left[  -\frac{\partial ^{2}}{\partial \beta_{1,k}^{2}}-\frac{3^k-5}{2}  \cot \beta_{1,k} +  \frac{B_{n_{1,k}}-\frac{1}{4}}{ \cos ^{2}\beta_{1,k}} +  \right.\right.\right.\nonumber\\
  & & \left.\left.\left.....    \right.\right.\right.\nonumber\\ 
 & & \left.\left.\left.
 \left[  -\frac{\partial ^{2}}{\partial \beta_{\ell,k-m} ^{2}}   -\frac{3^k-3^{m} -2 - 2  \ell}{2} \cot  \beta_{\ell,k-m} \frac{\partial }{\partial \beta_{\ell,k-m} } +  \frac{B_{n_{\ell,k-m}}-\frac{1}{4}}{ \cos ^{2}\beta_{\ell,k-m}} + 
  \right.\right.\right.\right. \nonumber\\  
 & & \left.\left.\left.\left.   ....   \right.\right.\right.\right.\nonumber\\ 
& &  \left.\left.\left.\left. \left[  -\frac{\partial ^{2}}{\partial \beta_{3^{k-1}-2,1} ^{2}}   - \cot  \beta_{3^{k-1}-2,1} \frac{\partial }{\partial \beta_{3^{k-1}-2,1} } + 
 \frac{B_{n_{3^{k-1}-2,1}}-\frac{1}{4}}{ \cos^{2}\beta_{1,3^{k-1}-2,1}}  
     \right.\right.\right.\right.\right. \nonumber\\
    & &  \left.\left.\left.\left.\left.
  + \frac{1}{\sin ^{2}\beta_{3^{k-1}-2,1}} \left[  -\frac{\partial ^{2}}{\partial \beta_{3^{k-1}-1,1}^{2}} +
 \frac{B_{n_{3^{k-1}-1,1}}-\frac{1}{4}}{ \cos^{2}\beta_{3^{k-1}-1,1}}+  \frac{B_{n_{3^{k-1},1}}-\frac{1}{4}}{ \sin^{2}\beta_{3^{k-1}-1,1}} \right]\right]\right] \right]\right] \right. \nonumber\\
& & \left.       -E_{ V_k({\bf n}) } \right\} \chi_{ V_k({\bf n}) }(r,\alpha,\hat{ V}_k({\bf \beta}))
 =0  \ .  \label{3ca-49f}
\end{eqnarray}%

 This equation is solved completely in the Appendix where   the  calculations are reported.

\section{Conclusions}

In this work, we have studied the exact solvability of a particular quantum  system of $N$ equal mass particles with $N=3^k \ (k \geq 2)$, confined in an harmonic field. In this  system, the particles are clustered in clusters of 3 particles. The interaction between the particles are governed by two-body Calogero potentials inside each cluster and with several many-body potentials. The number of these potentials increases with the number of the particles.
To illustrate the procedure for solving this quantum system, the particular case of 9 particles is studied $ (k=2)$ and solved
exactly. Namely the regular eigensolutions and the corresponding eigenenergies of the stationary Schr\"odinger equation are derived explicitly. The general case of $N=3^k$ particles is also studied. Thanks to some successive appropriate  coordinates transformations, the problem becomes separable, then the full solutions are explicitly derived, namely the eigenwave functions and the energy spectrum of the corresponding Schr\"odinger equation.
Having obtained the exact solution of this $N$-body particular quantum problem, with $N=3^k (k \geq 2)$, it appears that a similar$N$-body problem with $N=2^k (k \geq 2)$ particles  is also exactly solvable. In this case the particles are arranged  in clusters of 2-particles each, and interacting via a two-body Calogero potential inside each cluster, and with other many-body forces involved in the whole set of  interactions. 
Other solvable $N$-body problems may be obtained by replacing the confining harmonic term $\sum_{i=1}^N
\omega^2 x_i^2$ in the Hamiltonians considered in this paper, by an attractive "Coulomb-type" potential
$ -\alpha/\sqrt{\sum_{i=1}^N x_i^2} \ (\alpha > 0),$ giving rise to both bound states with negative discrete spectrum and scattering states with positive continuous spectrum.

{\bf Acknowledgements } We thank  Dr. R.J. Lombard for fruitful discussions.
One of us (A.B.) is  very grateful to the Theory Group of the IPN Orsay for its kind hospitality
and to the Mentouri  University of Constantine  for financial support.

\appendix
\section{Appendix}

In the whole section we consider values $ k \ne 1$. 
The  Hamiltonian Eq.(\ref{3ca-49f}) may be mapped to
 the problem of one particle in the space of dimension $(3^k+1)/2$ with a non central potential of the form 
\begin{eqnarray}
V(r,\alpha, \hat{ V}_k({\bf \beta}))&=& g(r)+\frac{1}{r^{2} \sin ^{2}\alpha } \left[ \sum_{m=0}^{k-1}
 \sum_{\ell=1}^{3^m}...  (1-\delta_{m,k-1} \delta_{\ell,3^{k-1}}) \left[ f_{\ell,k-m}(\beta_{\ell,k-m})     \right.\right. \nonumber\\ 
& &\left.\left. +  (1-\delta_{m,k-1} \delta_{\ell,3^{k-1}-1})  \frac{1}{\sin^2( \beta_{\ell,k-m})} \left[...
.....  \left[  f_{3^{k-1}-1,1}( \beta_{3^{k-1}-1,1}) \right]\right]\right] \right]. \nonumber
\end{eqnarray}
The problem becomes then  separable in the $(3^k+1)/2$  variables $\{r,\alpha,V_k({\bf \beta})\}$.
For the sake of simplicity we introduce the set :  
\begin{equation}
Z_{\ell,k-m}({\bf \Lambda,n}) \equiv  \hat{W}_{\ell,k-m}({\bf \Lambda}) \cup W_{\ell,k-m}({\bf n}) \ ,
\end{equation}
which is the union of $\hat{W}_{\ell,k-m}({\bf \Lambda})$ and $W_{\ell,k-m}({\bf n})$, and  which includes $3^k-2$ quantum numbers.  More precisely we have : 

\begin{eqnarray}
Z_{\ell,k-m}({\bf \Lambda,n}) & \equiv & \{\Lambda_{3^{k-1}-1,1},\Lambda_{3^{k-1}-2,1},...,\Lambda_{2,1},\Lambda_{1,1}, \Lambda_{3^{k-2},2},...,\Lambda_{1,2},...,   \nonumber\\
& &  \Lambda_{3^m,k-m},\Lambda_{3^{m}-1,k-m},....,\Lambda_{\ell+1,k-m},\Lambda_{\ell,k-m}, \nonumber\\
& & n_{3^{k-1},1},n_{3^{k-1}-2,1},...,n_{2,1},n_{1,1}, n_{3^{k-2},2},...,n_{1,2},...,   \nonumber\\
& &  n_{3^m,k-m},n_{3^{m}-1,k-m},....,n_{\ell+1,k-m},n_{\ell,k-m}\} 
\end{eqnarray}

To find the solution we factorize the wave function as follows :
\begin{eqnarray}
& & \chi_{n_r,n_{\alpha},Z_{1,k}({\bf \Lambda,n})}(r,\alpha,\hat{V}_k({\bf \beta})) =\frac{ F_{n_r,n_{\alpha},
Z_{1,k}({\bf \Lambda,n})}(r)}{r^{(3^{k}-1)/4}} \ \ 
\frac{ G_{n_{\alpha},Z_{1,k}({\bf \Lambda,n})}(\alpha)}{(\sin \alpha)^{(3^{k}-3)/4}} \nonumber\\
& & \times   \prod_{m=0}^{k-1} \prod_{\ell=1}^{3^{m}} (1-\delta_{m,k-1} \delta_{\ell,3^{k-1}}) 
 \frac{Q_{Z_{\ell,k-m}({\bf \Lambda,n})}(\beta_{\ell,k-m})}{(\sin \beta_{\ell,k-m})^{(3^{k}-3^m-2-2 \ell)/4}}, \quad
 k \geq 1 \  .  \label{3ca4-66} 
\end{eqnarray}
Accordingly,  equation (\ref{3ca-49f}) separates into $(3^k+1)/2$      decoupled  differential equations. The first one reads  : 
\begin{eqnarray}
& & \left(-\frac{\partial ^{2}}{\partial \beta_{3^{k-1}-1,1}^{2}}+ \frac{B_{n_{3^{k-1}-1,1}}-\frac{1}{4}}{ \cos^{2}\beta_{3^{k-1}-1,1}}+  \frac{B_{n_{3^{k-1},1}}-\frac{1}{4}}{ \sin^{2}\beta_{3^{k-1}-1,1}} \right. \nonumber\\
 & &  \left. - E_{\Lambda_{3^{k-1}-1,1},n_{3^{k-1},1},n_{3^{k-1}-1,1}}  \right)
  Q_{\Lambda_{3^{k-1}-1,1},n_{3^{k-1},1},n_{3^{k-1}-1,1} }(\beta_{1,3^{k-1}-1,1}) = 0 \ .
 \quad  \label{3ca-7f1}
\end{eqnarray} 
We remind that $B_{n_{\ell,k-m}},b_{n_{\ell,k-m}} $ are given by equations (\ref{3cb-22-bisp},\ref{3cb-22-terp}).
We have : 
\begin{eqnarray}
 E_{\Lambda_{3^{k-1}-1,1},n_{3^{k-1},1},n_{3^{k-1}-1,1}}& = & \epsilon_{\Lambda_{3^{k-1}-1,1},n_{3^{k-1},1},n_{3^{k-1}-1,1}}^2  \nonumber\\
\epsilon_{\Lambda_{3^{k-1}-1,1},n_{3^{k-1},1},n_{3^{k-1}-1,1}}  & = &2 \Lambda_{3^{k-1}-1,1} + 1 +b_{n_{3^{k-1},1}}+b_{n_{3^{k-1}-1,1}} \ .
\end{eqnarray} 
The eigensolution reads :
\begin{eqnarray}
 Q_{ \Lambda_{3^{k-1}-1,1},V_{k}({\bf n}) }(\beta_{3^{k-1}-1,1}) & = & 
 (\sin  \beta_{3^{k-1}-1,1} )^{\frac{1}{2}+ b_{3^{k-1},1}} \ 
  (\cos \beta_{3^{k-1}-1,1} )^{\frac{1}{2}+ b_{3^{k-1}-1,1}} \nonumber\\ 
& \times &  P_{\Lambda_{3^{k-1}-1,1}}^{( b_{3^{k-1},1}, b_{3^{k-1}-1,1})}(\cos 2 \beta_{3^{k-1}-1,1}) \ .
\end{eqnarray}
The next equation reads,
\begin{eqnarray}
& & \left(  -\frac{\partial ^{2}}{\partial \beta_{3^{k-1}-2,1}^{2}}    + 
 \frac{B_{n_{3^{k-1}-2,1}}-\frac{1}{4}}{ \cos^{2}\beta_{3^{k-1}-2,1}}  +
  \frac{E_{Z_{3^{k-1}-1,1}({\bf \Lambda,n})}  -\frac{1}{4}}{ \sin^{2}\beta_{3^{k-1}-2,1}} \right. \nonumber\\
& & \left. -  E_{Z_{3^{k-1}-2,1}({\bf \Lambda,n})}  \right)    
   Q_{Z_{3^{k-1}-2,1}({\bf \Lambda,n})}(\beta_{3^{k-1}-2,1} ) = 0 \ ,
 \quad  \label{3ca-7f2}
\end{eqnarray}%
 taking into account  :
\beq 
 E_{\Lambda_{3^{k-1}-1,1},n_{3^{k-1},1},n_{3^{k-1}-1,1}} \equiv  E_{Z_{3^{k-1}-1,1}({\bf \Lambda,n})} \ .
\eeq
We have : 
\begin{eqnarray}
  E_{Z_{3^{k-1}-2,1}({\bf \Lambda,n})} & = &   \epsilon_{Z_{3^{k-1}-2,1}({\bf \Lambda,n})}^2  
     \nonumber\\
 \epsilon_{Z_{3^{k-1}-2,1}({\bf \Lambda,n})}   & = &2 \Lambda_{3^{k-1}-2,1} + 2 \Lambda_{3^{k-1}-1,1} 
  +  2 +b_{n_{3^{k-1}-2,1}}+b_{n_{3^{k-1}-1,1}} + b_{n_{3^{k-1},1}} \ . \nonumber
\end{eqnarray} 
The corresponding eigensolution is : 
\begin{eqnarray}
 & &   Q_{Z_{3^{k-1}-2,1}({\bf \Lambda,n})}
 =(\sin  \beta_{3^{k-1}-2,1} )^{\frac{1}{2}+\epsilon_{Z_{3^{k-1}-1,1}({\bf \Lambda,n})}}
 \nonumber\\
 & & \times \   (\cos \beta_{3^{k-1}-2,1} )^{\frac{1}{2}+ b_{n_{3^{k-1}-2,1}}}
  \times   P_{\Lambda_{3^{k-1}-1,1}}^{(\epsilon_{Z_{3^{k-1}-1,1}({\bf \Lambda,n})},
    b_{n_{3^{k-1}-2,1}})}(\cos 2 \beta_{3^{k-1}-2,1}) \nonumber .
\end{eqnarray}
 The procedure is followed, and at the   $m^{\rm th}$ step,  we obtain : 
\begin{eqnarray}
& & \left( -\frac{\partial ^{2}}{\partial \beta_{\ell,k-m}^{2}}    + 
 \frac{B_{n_{\ell,k-m}}-\frac{1}{4}}{ \cos ^{2}\beta_{\ell,k-m}} +  \frac{ E_{Z_{\ell+1,k-m}({\bf \Lambda,n})} 
 -\frac{1}{4}}{ \sin^{2}\beta_{\ell,k-m} } \right. \nonumber\\
  & & \left. - E_{Z_{\ell,k-m}({\bf \Lambda,n})}  \right) 
 Q_{Z_{\ell,k-m}({\bf \Lambda,n})}( \beta_{\ell,k-m})=0  \ .
\end{eqnarray}
with
\begin{equation}
  E_{Z_{\ell,k-m}({\bf \Lambda,n})}=\epsilon_{Z_{\ell,k-m}({\bf \Lambda,n})}^2 \ .
 \end{equation} 
 We have
\begin{eqnarray}
\epsilon_{Z_{\ell,k-m}({\bf \Lambda,n})} &= & (1-\delta_{m,k-1}) \left(  \sum_{i=m+1}^{k-1} \sum_{j=1}^{3^{i}}
(2 \Lambda_{j,k-i}+1) (1-\delta_{i,k-1} \delta_{j,3^i})  + b_{j,k-i}  \right) \nonumber\\
&+& \sum_{j=\ell}^{3^m} [ (2 \Lambda_{j,k-m}+1)
(1-\delta_{m,k-1} \delta_{j,3^m})+ b_{j,k-m}  ] \ .
\end{eqnarray} 
 The eigensolution reads 
\begin{eqnarray}
 Q_{Z_{\ell,k-m}({\bf \Lambda,n})}( \beta_{\ell,k-m})& = & (\sin \beta_{\ell,k-m} )^{ \epsilon_{Z_{\ell+1,k-m}({\bf \Lambda,n})}+\frac{1}{2}}(\cos \beta_{\ell,k-m} )^{b_{\ell,k-m}+
\frac{1}{2}} \nonumber\\
& \times & P_{\Lambda_{k-m}}^{(\epsilon_{Z_{\ell+1,k-m}({\bf \Lambda,n})},
 b_{n_{\ell,k-m}})}(\cos 2\beta_{\ell,k-m} ), \label{thetapaf}\\
& & \quad 0 \leq \beta_{\ell,k-m} \leq \frac{\pi }{2},\quad \Lambda_{k-m}=0,1,2,... \ . \nonumber
\end{eqnarray}
The  equation concerning the angular variable $\beta_{1,k}$ is written as 
\begin{equation}
\left( -\frac{\partial ^{2}}{\partial \beta_{1,k}^{2}}    + 
 \frac{B_{n_{1,k}}-\frac{1}{4}}{ \cos ^{2}\beta_{1,k}} +  \frac{ E_{Z_{2,k}({\bf \Lambda,n})} -\frac{1}{4}}{ \sin^{2}\beta_{1,k} }-  E_{Z_{1,k}({\bf \Lambda,n})} \right) 
 Q_{Z_{1,k}({\bf \Lambda,n})}( \beta_{1,k})=0 
\label{eqf}
\end{equation}
with
\begin{equation}
 \epsilon_{Z_{1,k}({\bf \Lambda,n})}=  \left(  \sum_{i=0}^{k-1} \sum_{j=1}^{3^{i}}
(2 \Lambda_{j,k-i}+1) (1-\delta_{i,k-1} \delta_{j,3^i})  + b_{j,k-i}  \right) \ .
\end{equation} 
The latter equation (\ref{eqf}) includes $3^k-2$ quantum numbers. The two last equations are :
\begin{equation}
\left( -\frac{\partial ^{2}}{\partial \alpha ^{2}}   + \frac{ E_{Z_{1,k}({\bf \Lambda,n})}  -\frac{1}{4}}{
{ \sin^{2}\alpha} }-  E_{n_{\alpha},Z_{1,k}({\bf \Lambda,n})} \right) 
G_{n_{\alpha},Z_{1,k}({\bf \Lambda,n})}(\alpha) =0
\end{equation}
with
\begin{eqnarray}
E_{n_{\alpha},Z_{1,k}({\bf \Lambda,n})} & = & \epsilon_{n_{\alpha},Z_{1,k}({\bf \Lambda,n})}^2 \nonumber\\
 \epsilon_{n_{\alpha},Z_{1,k}({\bf \Lambda,n})} & = & \left(  \sum_{i=0}^{k-1} \sum_{j=1}^{3^{i}}
(2 \Lambda_{j,k-i}+1) (1-\delta_{i,k-1} \delta_{j,3^i})  + b_{j,k-i} \right) + n_{\alpha}+\frac{1}{2}
\end{eqnarray}

and the radial equation 
\begin{equation}
\left( -\frac{\partial ^{2}}{\partial r^{2}}
+\omega ^{2}r^{2}+\frac{\mu+E_{n_{\alpha},Z_{1,k}({\bf \Lambda,n})}-\frac{1}{4} }{r^{2}}  -
E_{n_r,n_{\alpha},Z_{1,k}({\bf \Lambda,n})} \right) F_{n_r,n_{\alpha},Z_{1,k}({\bf \Lambda,n})}(r)=0
\end{equation}
with
\begin{equation}
\frac{E_{n_r,n_{\alpha},Z_{1,k}({\bf \Lambda,n})}}{2 \omega}= \left \{ 2 n_r + 1 +\sqrt{ \mu + 
 \left[  \left(\sum_{i=0}^{k-1} \sum_{j=1}^{3^{i}} (2 \Lambda_{j,k-i}+1)(1-\delta_{i,k-1} \delta_{j,3^i})     + b_{j,k-i}  \right) + 
n_{\alpha}+\frac{1}{2} \right]^2 }
 \right \} 
\end{equation} 
Setting

\begin{equation}
d_{\alpha}=\epsilon_{Z_{1,k}({\bf \Lambda,n})} \ ,  \quad  
\kappa^2=\mu+E_{n_{\alpha},Z_{1,k}({\bf \Lambda,n})} \ ,
\end{equation}
and taking into account the equations (\ref{3ca-30}),(\ref{3ca37}) and (\ref{thetapaf}), the final solution, Eq.(\ref{3ca4-66}), reads :
\begin{eqnarray}
& &\chi_{n_r,n_{\alpha},Z_{1,k}({\bf \Lambda,n})}(r,\alpha,\hat{V}_k({\bf \beta})) = 
r^{\kappa-(3^k-3)/4} L_{n_r}^{\kappa}(\omega r^2) \exp \left( - \frac{\omega r^2}{2} \right)
\ \sin(\alpha)^{d_{\alpha}-(3^k-5)/4} \ C_{n_{\alpha}}^{(d_{\alpha}+1/2)}(\cos \alpha) \nonumber\\
& & \prod_{m=0}^{k-1}  \prod_{\ell=1}^{3^m} (1-\delta_{m,k-1} \delta_{\ell,3^m})
 (\sin \beta_{\ell,k-m} )^{ \epsilon_{Z_{\ell,k-m}({\bf \Lambda,n})}+
 1 -(3^k-3^m-2 \ell)/4} (\cos \beta_{k-m} )^{b_{\ell,k-m}+\frac{1}{2}} \nonumber\\
& & \times  P_{\Lambda_{k-m}}^{(\epsilon_{Z_{\ell,k-m}({\bf \Lambda,n})},b_{\ell,k-m})}(\cos 2\beta_{k-m} ) \ . \nonumber
\end{eqnarray}

The general solution $\Psi_{n_r,n_{\alpha},Z_{1,k}({\bf \Lambda,n})}(V_{k}({\bf r}),V_{k}({\bf \varphi}))$, Eq.(\ref{genpsi}), in its symmetric form,  reads :

\begin{eqnarray}
& & \Psi_{n_r,n_{\alpha},Z_{1,k}({\bf \Lambda,n})}(V_{k}({\bf r}),V_{k}({\bf \varphi}))   = 
r^{\kappa+1-3^{k}/2} L_{n_r}^{\kappa}(\omega r^2) \exp \left( - \frac{\omega r^2}{2} \right) \  \sin(\alpha)^{d_{\alpha}+3/2-3^{k}/2} \ C_{n_{\alpha}}^{(d_{\alpha}+1/2)}(\cos \alpha) \nonumber\\
& & \prod_{m=0}^{k-1}  \prod_{\ell=1}^{3^{m}} (1-\delta_{m,k-1} \delta_{\ell,3^m})
 (\sin \beta_{\ell,k-m} )^{ \epsilon_{Z_{\ell,k-m}({\bf \Lambda,n})} -(3^k-3^m-2 \ell-2)/2} (\cos \beta_{k-m} )^{b_{\ell,k-m}} \nonumber\\
& & \times  P_{\Lambda_{k-m}}^{( \epsilon_{Z_{\ell,k-m}({\bf \Lambda,n})},b_{\ell,k-m})}(\cos 2\beta_{k-m} ) 
 \ \prod_{m=1}^k \prod_{\ell=1}^{3^{k-m}} \vert \sin 3\varphi_{\ell,m} \vert^{\frac{1}{2}+ a_{\ell,m}} C_{n_{\ell,m}}^{(\frac{1}{2}+a_{\ell,m})}(\cos 3\varphi_{\ell,m} ) \ ,
\end{eqnarray}
\begin{eqnarray}
&& (\forall \ell)  (\forall m) \quad \Lambda_{\ell,k-m} =0,1,2,...  \quad 1 \leq \ell \leq 3^{k-m}, 1  \leq m \leq  k-1 \nonumber\\
 & &  (\forall \ell)  (\forall m) \quad n_{\ell,k-m} =0,1,2,...  \quad 1 \leq \ell \leq 3^{k-m}, 1  \leq m \leq  k-1 \nonumber\\
  & &  (\forall \ell)  (\forall m) \quad  0 \leq \beta_{\ell,k-m} \leq \frac{\pi}{2}, \quad 1 \leq \ell \leq 3^{k-m}, 1  \leq m \leq  k-1 \nonumber\\
  & & 0 \leq \alpha \leq \pi \quad 0 \leq r < \infty \nonumber\\
   & &  (\forall \ell)  (\forall m) \quad  0 \leq \varphi_{\ell,k-m}  \leq \frac{\pi}{3},  \quad 1 \leq \ell \leq 3^{k-m}, 1  \leq m \leq  k \nonumber\\
     & &  (\forall \ell)  (\forall m)  \quad  a_{\ell,m}   =  \frac{1}{2} \sqrt{1+2\lambda_{\ell,m} } \ ,  \quad 1 \leq \ell \leq 3^{k-m}, 1  \leq m \leq  k \ .
\label{FIN}
\end{eqnarray}
Note that $ \Lambda_{3^k,1} $ does not appear  in Eqs.(\ref{FIN}).

\end{document}